\documentclass{aa}  
\usepackage{graphicx}
\usepackage{txfonts}
\usepackage{amsmath}
\usepackage{graphicx}
\usepackage[colorlinks=true, allcolors=blue]{hyperref}
\usepackage{systeme}
\usepackage[ruled,noline,noend,linesnumbered]{algorithm2e}
\usepackage{listings}
\usepackage{booktabs} 
\usepackage{multirow}
\usepackage{subcaption}
\usepackage{float}
\usepackage{soul}
\usepackage{etoolbox} % For toggles
\usepackage{xcolor}
\begin{document}

\title{Improving the discovery of near-Earth objects with machine-learning methods.}
%\thanks{Accepted for publication in \textit{Astronomy \& Astrophysics}. DOI: \href{https://doi.org/10.1051/0004-6361/202554311}{10.1051/0004-6361/202554311}}
%\thanks{Accepted for publication in 	extit{Astronomy \& Astrophysics}}

\author{Peter Vere\v{s}\inst{1}
        \thanks{These authors contributed equally to this work.}
      \and Richard Cloete\inst{1}\footnotemark[1]
      \and Matthew J. Payne\inst{1}
      \and Abraham Loeb\inst{1}
}

\institute{Harvard-Smithsonian Center for Astrophysics, 60 Garden St., MS 15, Cambridge, MA 02138, USA \\
\email{pveres@cfa.harvard.edu}
}

\abstract
% context
{We present a comprehensive analysis of the \textit{digest2} parameters for candidates of the Near-Earth Object Confirmation Page (NEOCP) that were reported between 2019 and 2024. Our study proposes methods for significantly reducing the inclusion of non-NEO objects on the NEOCP. Despite the substantial increase in near-Earth object (NEO) discoveries in recent years, only about half of the NEOCP candidates are ultimately confirmed as NEOs. Therefore, much observing time is spent following up on non-NEOs. Furthermore, approximately 11\% of the candidates remain unconfirmed because the follow-up observations are insufficient. These are nearly 600 cases per year.}
% aims
{To reduce false positives and minimize wasted resources on non-NEOs, we refine the posting criteria for NEOCP based on a detailed analysis of all \textit{digest2} scores.}
% methods
{We investigated 30 distinct \textit{digest2} parameter categories for candidates that were confirmed as NEOs and non-NEOs. From this analysis, we derived a filtering mechanism based on selected \textit{digest2} parameters that were able to exclude 20\% of the non-NEOs from the NEOCP while maintaining a minimal loss of true NEOs. We also investigated the application of four machine-learning (ML) techniques, that is, the gradient-boosting machine (GBM), the random forest (RF) classifier, the stochastic gradient descent (SGD) classifier, and neural networks (NN) to classify NEOCP candidates as NEOs or non-NEOs. Based on \textit{digest2} parameters as input, our ML models achieved a precision of approximately 95\% in distinguishing between NEOs and non-NEOs.}
% results 
{Combining the \textit{digest2} parameter filter with an ML-based classification model, we demonstrate a significant reduction in non-NEOs on the NEOCP that exceeds 80\%, while limiting the loss of NEO discovery tracklets to 5.5\%. Importantly, we show that most follow-up tracklets of initially misclassified NEOs are later correctly identified as NEOs. This effectively reduces the net loss of true NEOs to approximately 1\%.}
% conclusions 
{A greater purity of NEO candidates on the NEOCP would allow follow-up observers to allocate more resources to confirming high-priority objects. This would improve the overall observational efficiency and the confirmation rate of NEO discoveries. We suggest that our methods are used as part of the NEOCP pipeline.}


\maketitle

\section{Introduction}
\label{sec:intro}

Near-Earth objects (NEOs) are natural celestial bodies, including asteroids and comets, on heliocentric orbits with perihelion distances shorter than 1.3 AU. The interest in NEOs has grown significantly throughout the 20th century, driven by the realization that these objects have the potential to impact Earth~\citep{Fairchild07,Opik16,Opik76,Alvarez80,Morrison06}. Although the first NEO, 433 Eros, was discovered in 1898 through photographic plate observations, the systematic discovery and study of these objects did not gain momentum until much later. In 1998, the US Congress mandated NASA to discover 90\% of the NEOs that are larger than 1 km~\citep{Morrison92}; in 2005, this target was expanded to include NEOs larger than 140 meters (George E. Brown Act\footnote{\url{https://www.congress.gov/congressional-report/109th-congress/house-report/158/1}}). Together with advances in technology, in particular, the development of charge-coupled device (CCD) cameras, these legislative efforts sparked a dramatic growth in the number of known NEOs from only a few hundred in 1990 to more than 37,000 today. Dedicated offices at the national space agencies, such as NASA's Planetary Defense Coordination Office\footnote{\url{https://science.nasa.gov/planetary-science/programs/planetarydefense/}} and ESA's Planetary Defense Office in the Space Situational Awareness program\footnote{\url{https://www.esa.int/Space_Safety/Planetary_Defence}}, further reflect the international commitment to planetary defense. This is exemplified by evolving strategies and plans\footnote{\url{https://www.nasa.gov/wp-content/uploads/2023/06/nasa_-_planetary_defense_strategy_-_final-508.pdf},\url{https://www.nasa.gov/wp-content/uploads/2022/03/ostp-neo-strategy-action-plan-jun18.pdf}}.

The study of NEOs extends beyond planetary defense initiatives to include exploration, resource utilization, and understanding of the evolution of our Solar System. Recent examples include Hayabusa~\citep{Itokawa}, Hayabusa2~\citep{Hayabusa2}, and OSIRIS-REx~\citep{Lauretta17}, while the DART mission~\citep{Rivkin21} has successfully demonstrated the kinetic impactor technique for planetary defense. Upcoming missions such as Hera~\citep{Michel22}, Ramses~\citep{Ramses}, Tianwen-2~\citep{Cheng24}, and the commercial AstroForge venture~\citep{Astroforge} offer further opportunities to expand our understanding of these celestial bodies.

\begin{figure}[ht]
\includegraphics[width=\linewidth]{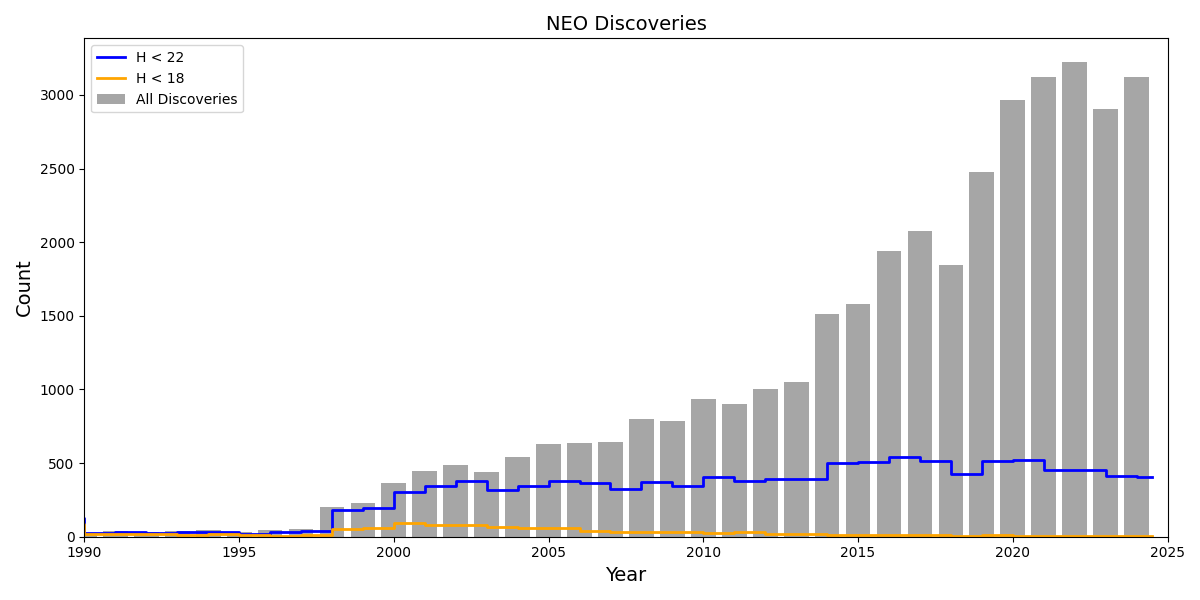}
    \caption{Number of NEO discoveries per year. Large NEOs with a size of > 1 km (H<18) and the NASA target (H<22) are displayed as well.}
    \label{fig:neocp_dist_stats}
\end{figure}

Despite significant progress and record-breaking NEO discovery rates in the past few years (Figure~\ref{fig:neocp_dist_stats}), challenges remain to meet the 2005 mandated objective of identifying 90\% of the NEOs that are larger than 140 meters. Current estimates suggest that less than 40\% of this population has been discovered~\citep{Grav23}, in contrast to the >90\% completion rate for the NEOs that are larger than one kilometer. NASA-funded surveys, such as the Catalina Sky Survey~\citep{Christensen12}, Pan-STARRS~\citep{Kaiser02}, and ATLAS~\citep{Tonry18}, have been instrumental in driving discoveries in the past 15 years. Upcoming initiatives, including the Rubin Observatory~\citep{jones2009solar,Veres17,Jones18}, which is set to begin operations in 2025, and the infrared space telescope NEO Surveyor~\citep{Grav20}, are expected to substantially enhance the detection capabilities and to fulfill the 140-meter discovery objective~\citep{Veres17,Mainzer23,Wagg24}.

The process of a NEO discovery relies heavily on the rapid reporting of potential candidates to the Minor Planet Center (MPC\footnote{\url{https://minorplanetcenter.net}}) Near-Earth Object Confirmation Page\footnote{\url{https://minorplanetcenter.net/iau/NEO/toconfirm_tabular.html}} (NEOCP). Follow-up observations allow a quick orbit determination and the formal designation of new NEOs. Central to this process is the use of a tool known as \textit{digest2} \citep{keys2019digest2}, which produces a quasi-probabilistic metric known as the \textit{digest2} score. This \textit{digest2} score is used to  evaluate the likelihood that a single tracklet, that is, a set of detections recorded over a short period, is a NEO. A specific score variant, the NEO noid digest2 score, is used to determine whether a tracklet is eligible for inclusion on the NEOCP. For this, the score must be 65 or greater.

Although the NEO noid \textit{digest2} score effectively classifies high-scoring candidates, its reliability diminishes below certain thresholds~\citep{keys2019digest2}. Many non-NEOs therefore appear on the NEOCP. In 2024, for example, roughly 6,000 candidates were posted, only about 55\% of which proved to be genuine NEOs, while 11\% remained unconfirmed. The high proportion of non-NEOs consumes valuable telescope time for follow-up observations. Moreover, previous work by \citet{Veres2018} suggested that most unconfirmed candidates may still be genuine NEOs. This underscores the need for better screening mechanisms.

To reduce false positives and to minimize wasting resources on non-NEOs, this study proposes refined posting criteria for NEOCP. We explore additional \textit{digest2} parameters and employ machine-learning techniques to improve the classification accuracy. The ultimately aim is to allocate follow-up efforts more efficiently and to improve the confirmation rate for true NEOs.

\section{NEOCP data}
\label{sec:Data}

\begin{figure}[ht]
\includegraphics[width=\linewidth]{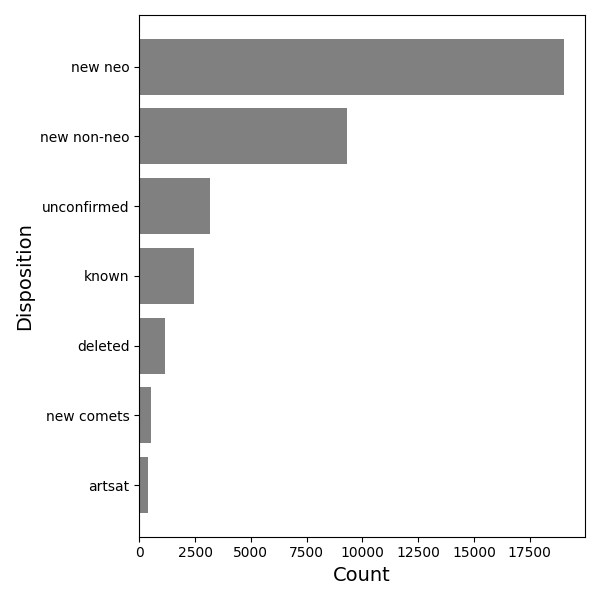}
    \caption{Final disposition of the NEOCP tracklets between March 14, 2019 and December 31, 2024.}
    \label{fig:neocp_final_disp}
\end{figure}

\begin{figure}[ht]
\includegraphics[width=\linewidth]{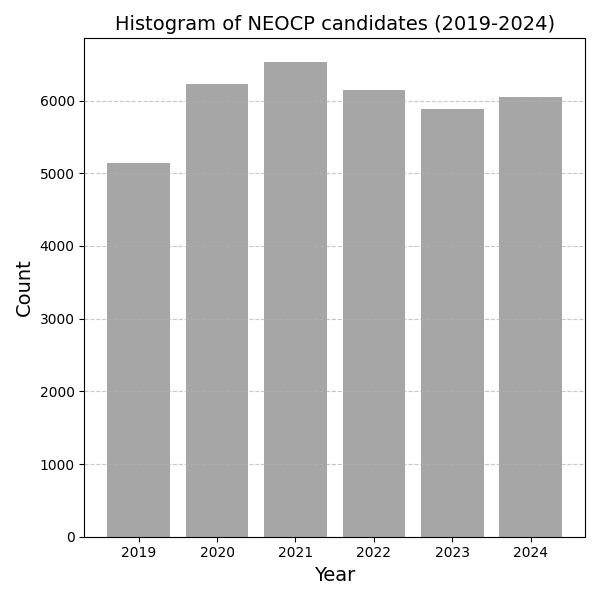}
    \caption{Number of NEOCP candidates posted per year.}
    \label{fig:neocp_cand_stats}
\end{figure}

\begin{figure}[ht]
\includegraphics[width=\linewidth]{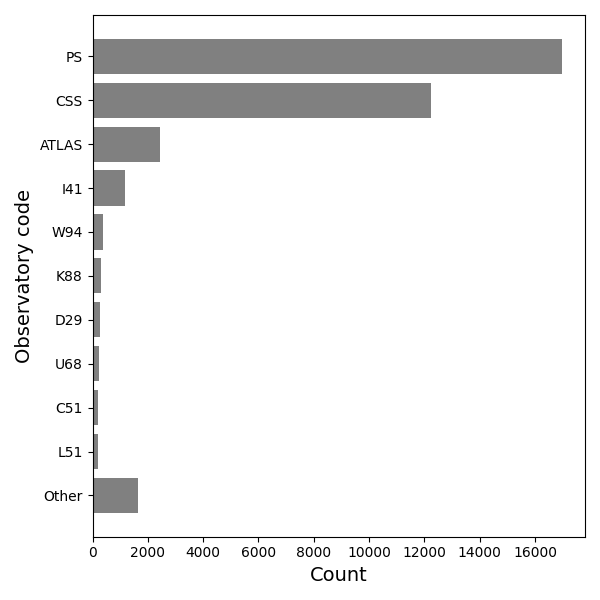}
    \caption{Most productive observatory codes on the NEOCP between 2019-2024. The ATLAS survey has four telescopes: T05, T08, W68, and M22. Pan-STARRS
(PS) has two telescopes: F51 and F52. The Catalina Sky Survey (CSS) has five telescopes:
G96, 703, I52, V00, and V06.}
    \label{fig:obscode_stats_neocp}
\end{figure}

We selected initial (discovery) tracklets for each object that appeared on NEOCP between March 14, 2019 and December 31, 2024. Although these six years included a record annual number of NEOCP candidates~ (Figure~\ref{fig:neocp_cand_stats}), the rate of new postings since 2019 was relatively steady compared to previous years (Figure~\ref{fig:neocp_dist_stats}). The vast majority of NEOCP candidates were reported by major NEO-dedicated surveys: Pan-STARRS, the Catalina Sky Survey, and ATLAS (Figure~\ref{fig:obscode_stats_neocp}). In total, the dataset contained 36,046 discovery tracklets. The final disposition categories for these NEOCP candidates are illustrated in Figure~\ref{fig:neocp_final_disp}. 

A separate section of the NEOCP, the Possible Comet Confirmation Page (PCCP)\footnote{\url{https://www.minorplanetcenter.net/iau/NEO/pccp_tabular.html}}, is used for objects exhibiting cometary activity (a coma and/or tail), which are posted regardless of their NEO \textit{digest2} scores. Because comets on the PCCP are recognized by their appearance rather than purely by the \textit{digest2} scoring, newly discovered comets were excluded from this study. In general, NEOCP objects can have a variety of final dispositions: 
\begin{itemize} 
    \item \emph{New NEO}: A genuine new NEO discovery that is eventually announced via an online publication (discovery MPEC). 
    \item \emph{New non-NEO}: A new designation for an object found to be a non-NEO. 
    \item \emph{Known}: A previously identified minor planet (including known NEOs observed at a subsequent apparition). 
    \item \emph{Artificial}: Approximately 1\% of the NEOCP objects turn out to be human-made satellites or debris. 
    \item \emph{Deleted}: Approximately 3\% of the postings are removed because they are deemed spurious (e.g., artifacts, stars, or noise). 
    \item \emph{Unconfirmed}: About 11\% of candidates remain unconfirmed (i.e., they receive insufficient follow-up) and are declared lost. They are ultimately placed in the isolated tracklet file for potential future recovery or orbit linkage. 
\end{itemize}

Consequently, the main subset of data that we analyzed comprises 19,019 new NEOs (including any previously unidentified second-apparition recoveries) and 9,306 new non-NEOs. Known objects were not considered, assuming that they were missed by internal MPC attribution mechanisms, but were quickly removed from NEOCP as known.

\section{Digest2}
\label{sec:digest2}

The \textit{digest2} code \citep{keys2019digest2} has been used by the NEO community for almost two decades. Implemented in the C programming language, it is publicly available online\footnote{\url{https://bitbucket.org/mpcdev/digest2}} and processes observations in the MPC1992 astrometric format. An improved version capable of reading ADES-XML astrometry is also publicly available \citep{Veres23}. Furthermore, a set of observations can be assessed by \textit{digest2} online on the MPC website\footnote{\url{https://minorplanetcenter.net/iau/NEO/PossNEO.html}}. 

The \textit{digest2} code generates a large set of unperturbed Sun-bound variant orbits that match the end-points of the input tracklet. It then classifies these orbits into distinct orbital categories based on perihelion distance, eccentricity, inclination, and absolute magnitude ($H$), and it calculates the frequency of orbits within each class. For example, if 90\% of the generated orbits correspond to NEO orbits, the resulting NEO score is 90. Thus, \textit{digest2} functions as a quasi-probabilistic classification tool.

\begin{figure}[ht]
   \includegraphics[width=\linewidth]{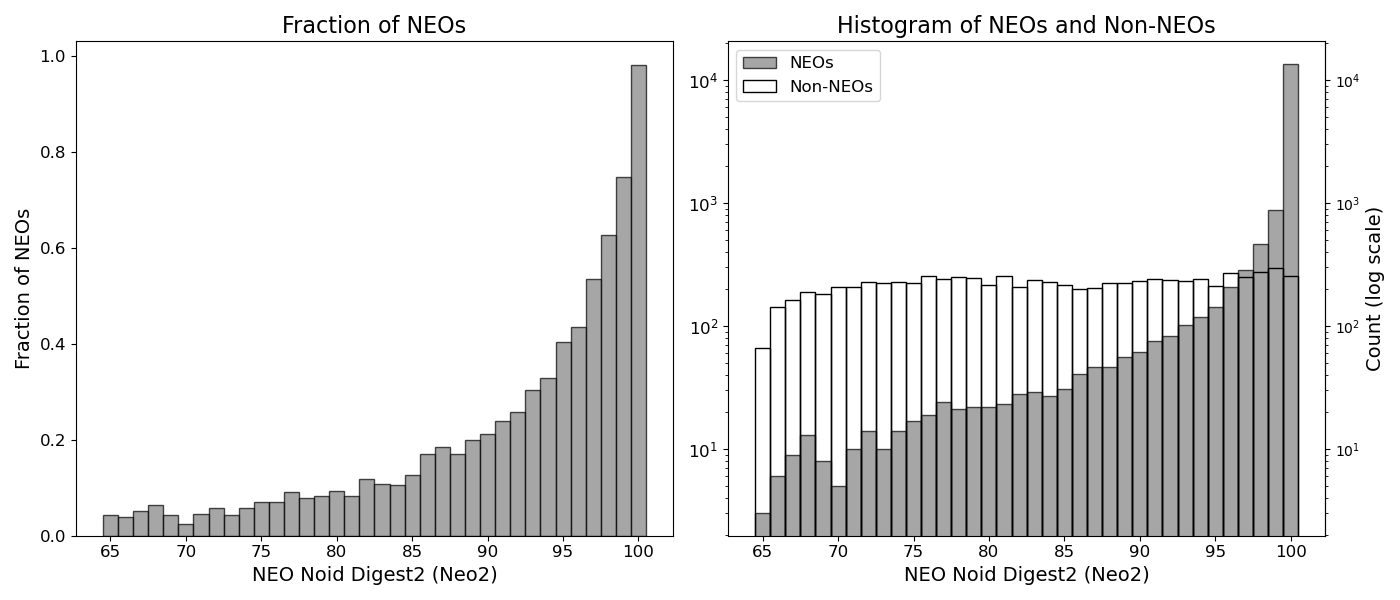}
    \caption{Fraction of NEOs on NEOCP in the NEO noid digest2 score (left) and the score of NEOs and non-NEOs on NEOCP in 2019-2024.}
    \label{fig:main_hist}
\end{figure}

Traditionally, only a single \texttt{digest2} value (the NEO noid score) was used to decide whether an object should be posted on the NEOCP. The noid variant of the score represents the likelihood that an object is a NEO relative to the undiscovered fraction of the Solar System population. The NEOCP posting threshold for the NEO noid score has historically been set at 65. Although this cutoff maximizes the number of NEO discoveries, it also results in a high fraction of non-NEOs being posted to the NEOCP. Figure~\ref{fig:main_hist} illustrates the sharp decline in the fraction of true NEOs as a function of the NEO noid \textit{digest2} score. In recent years, approximately 6{,}000 candidates have appeared on the NEOCP annually, and about 11\% remained unconfirmed.

\begin{figure}[ht]
   \includegraphics[width=\linewidth]{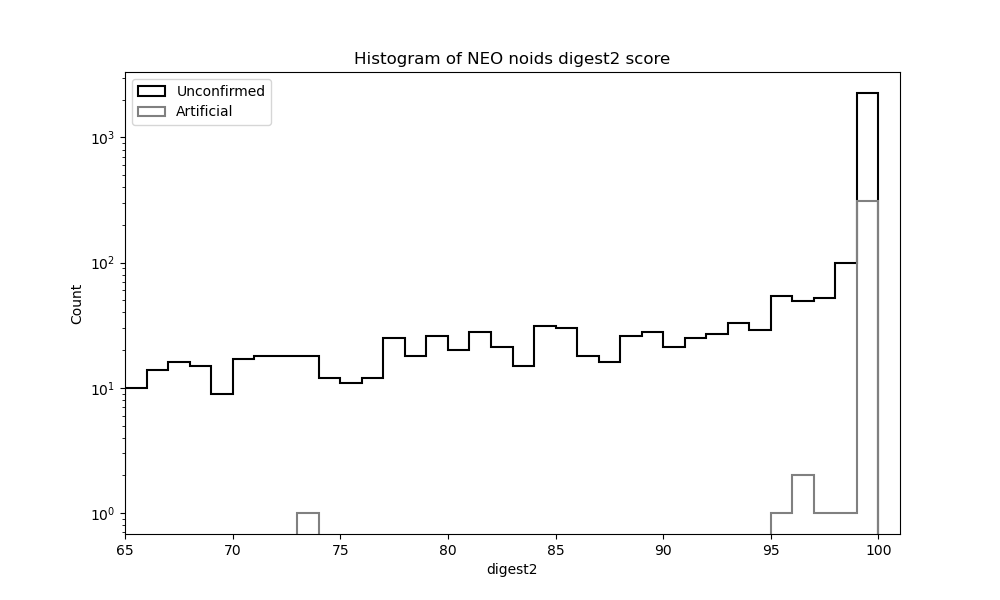}
    \caption{NEO noid digest2 score of 3,725 unconfirmed (dark) and artificial objects (light) on NEOCP in 2019-2024.}
    \label{fig:main_hist_unconfirmed}
\end{figure}

The typical NEO noid \textit{digest2} scores for unconfirmed and artificial objects on the NEOCP are shown in Figure~\ref{fig:main_hist_unconfirmed}. The score distribution for unconfirmed objects closely resembles that of all NEOCP candidates (Figure~\ref{fig:main_hist}), while artificial objects typically exhibit very high \textit{digest2} scores. These high scores are often due to their proximity to Earth and to their high apparent rates of motion.

Although the NEO noid score has become the canonical reference, \texttt{digest2} can generate up to 15 distinct scores for a single tracklet, each corresponding to a different orbit class (see Table~8 in~\citealt{keys2019digest2}). Each score is produced in two variants: the noid score, based on the undiscovered fraction of the Solar System population, and a raw score, based on the overall Solar System population. The orbital categories are as follows: interesting (Int) objects (high inclination and eccentricity), NEOs, Mars crossers (MC), Hungarias (Hun), Phocaeas (Pho), inner main belt asteroids (MBA1), Pallas family (Pal), Hansa family (Han), central main belt asteroids (MBA2), outer main belt asteroids (MBA3), Hildas (Hil), Jupiter trojans (Jtr), and Jupiter-family comets (JFC). Two additional classes, NEO18 (NEOs with $H<18$) and NEO22 (NEOs with $H<22$), are available, but are not used in this work.

\section{Digest2 analysis of NEO candidates}
\label{sec:digest2 filtering}

\begin{figure}[ht]
   \includegraphics[width=\linewidth]{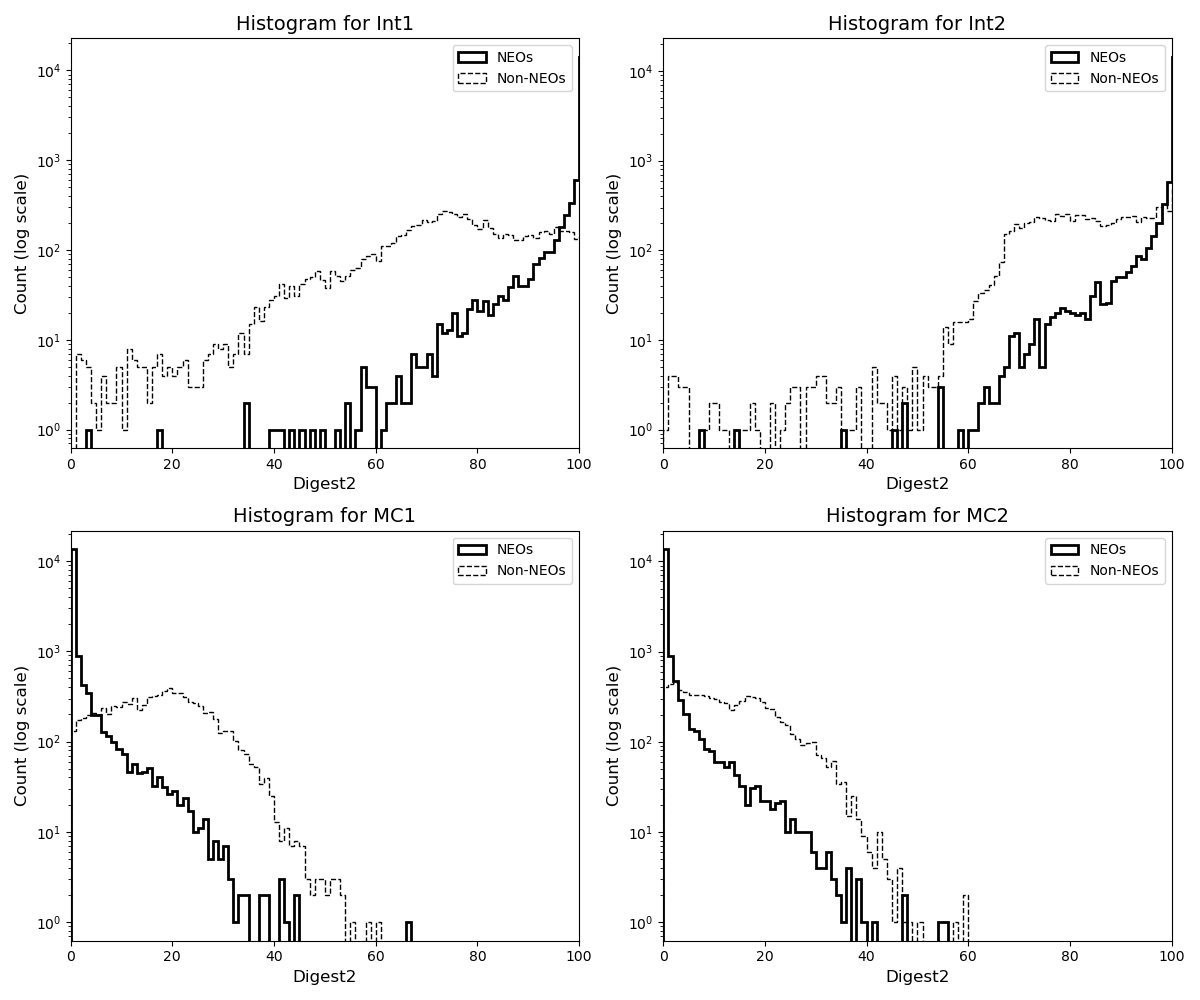}
   \includegraphics[width=\linewidth]{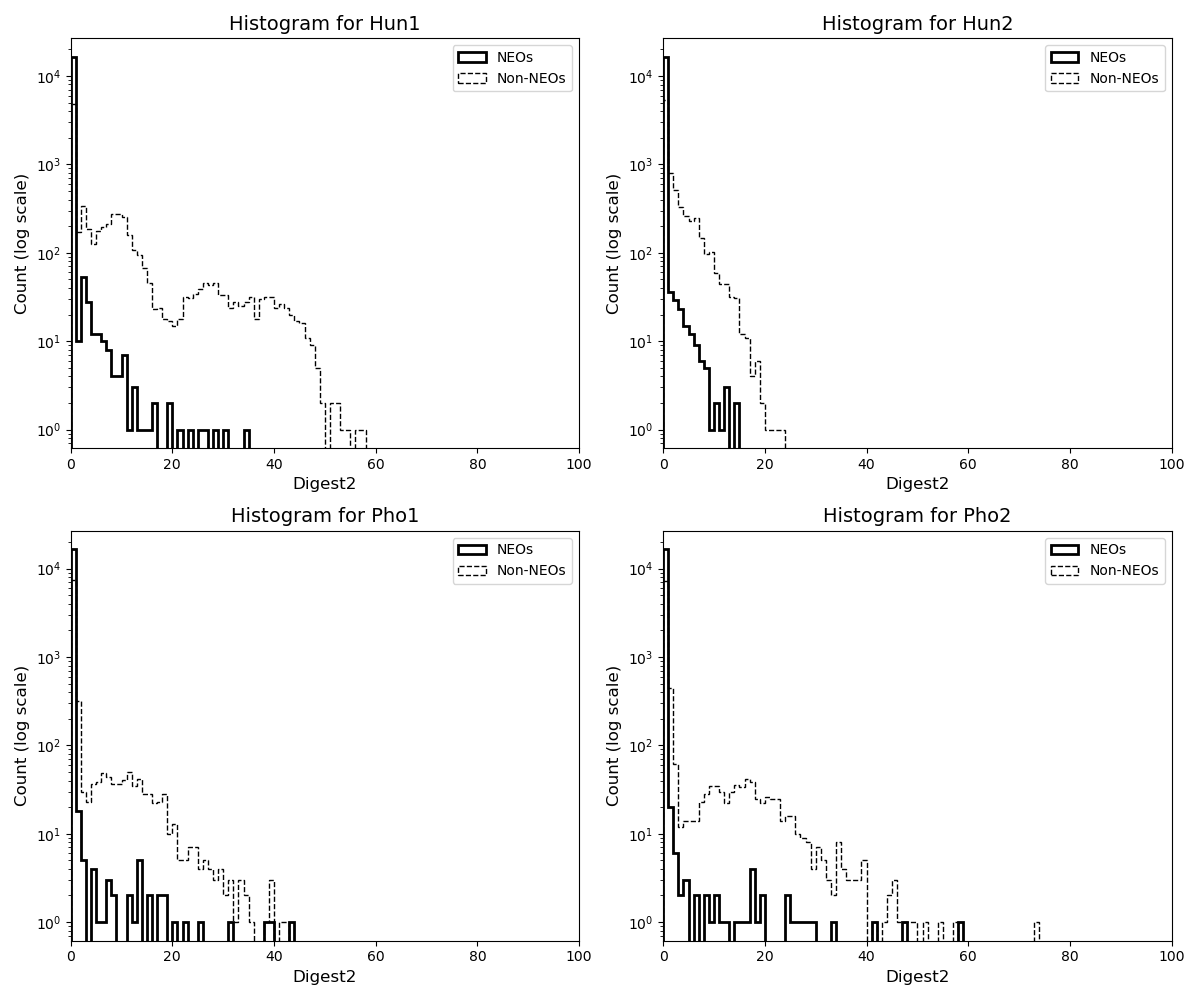}
    \caption{Histogram of the Int, MC and Hun and Pho digest2 scores for NEOs and non-NEOs. The number 1 denotes raw,  and the number 2 denotes noid.}
    \label{fig:int_mc_hun_pho}
\end{figure}

\begin{figure}[ht]
    \includegraphics[width=\linewidth]{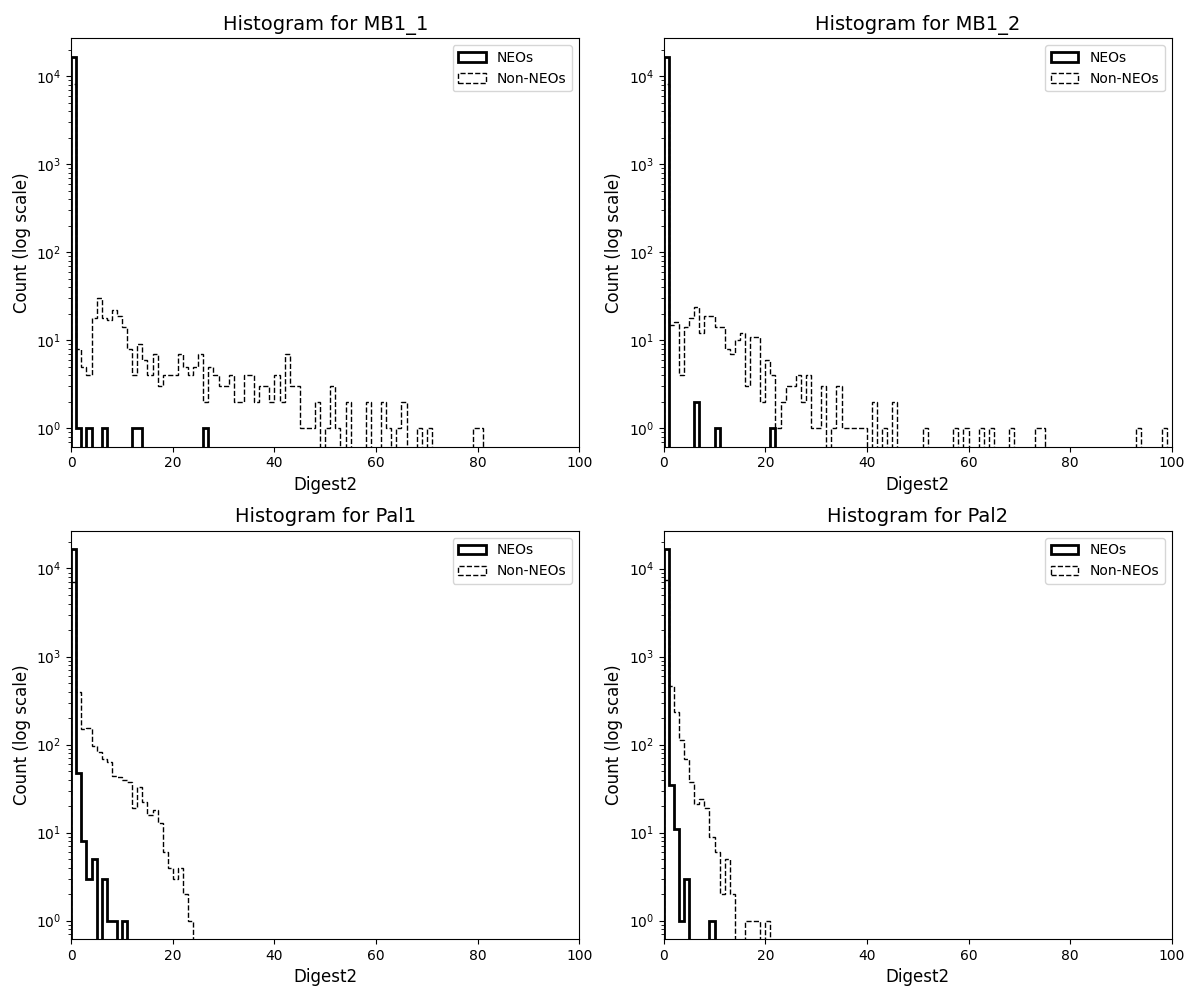}
\includegraphics[width=\linewidth]{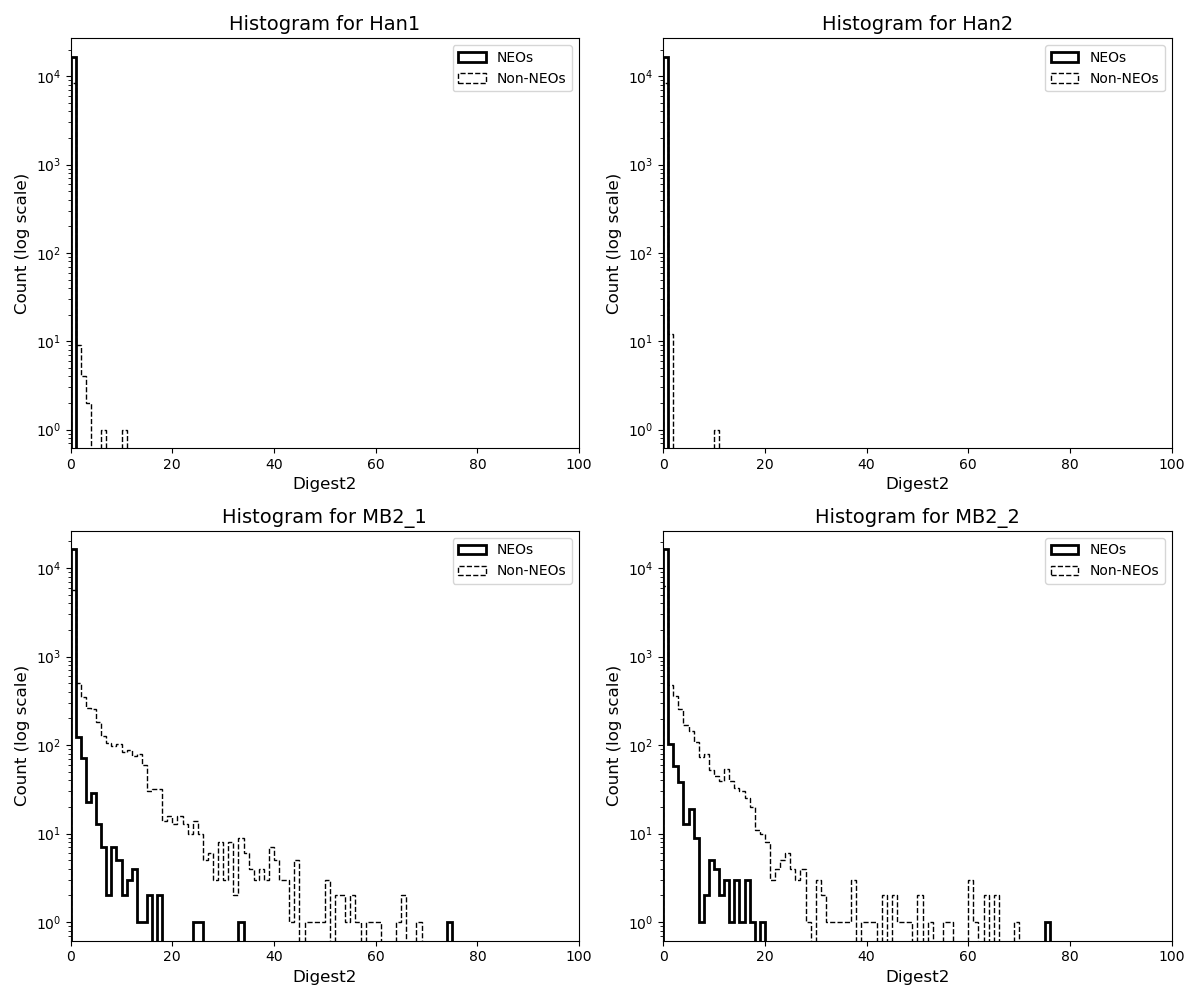}

    \caption{Histogram of the MBA, Pal, and Han and MBA2 digest2 scores for NEOs and non-NEOs. The number 1 denotes raw,  and the number 2 denotes noid.}
    \label{fig:mba_pal_han_mba2}
\end{figure}

\begin{figure}[ht]
\includegraphics[width=\linewidth]{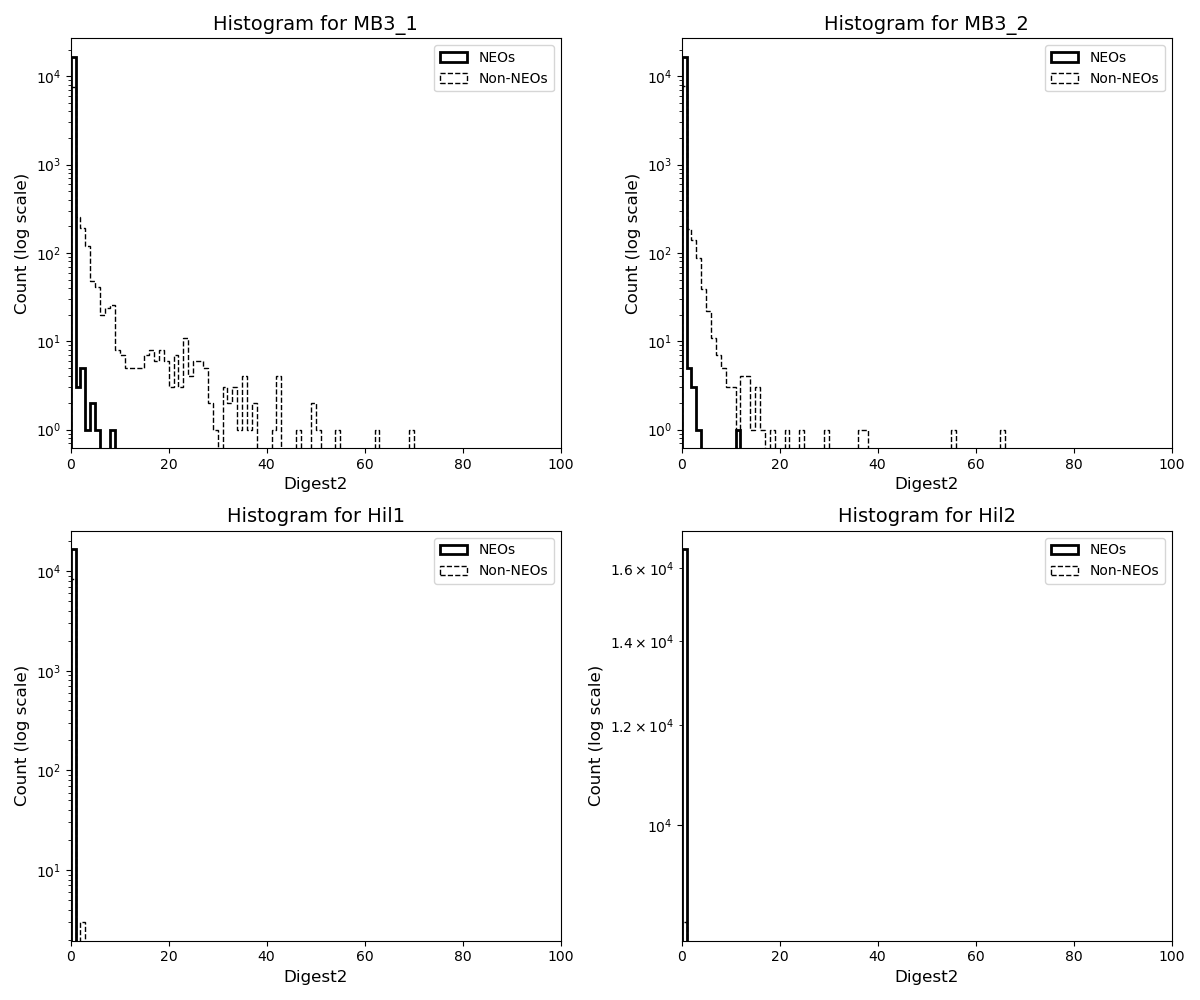}
\includegraphics[width=\linewidth]{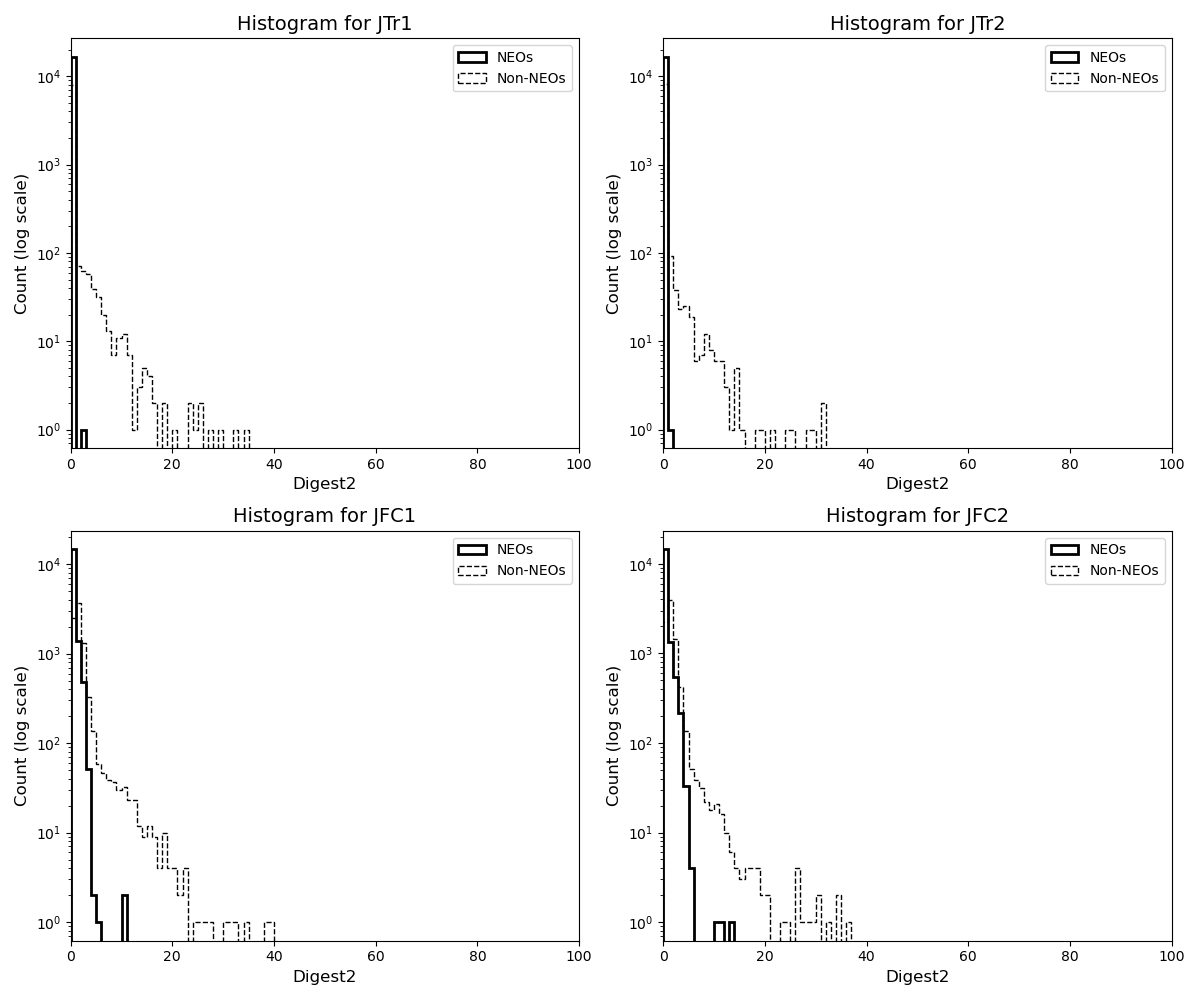}

    \caption{Histogram of Outer Main Belt (MBA), Hildas, Jupiter Trojans and Jupiter family comets digest2 scores for NEOs aThe number 1 denotes raw,  and the number 2 denotes noid.}
    \label{fig:mba3_hil_jtr_jfc}
\end{figure}

Using discovery tracklets from 2019--2024, we computed separate \texttt{digest2} scores for each of the 13 orbital categories in the raw and noid variants (the suffixes \emph{1} and \emph{2} in the figures, respectively). The histograms in Figures~\ref{fig:int_mc_hun_pho}, \ref{fig:mba_pal_han_mba2}, and \ref{fig:mba3_hil_jtr_jfc} illustrate how these scores compare for NEOs and non-NEOs in the categories. In some cases, the distinction is fairly strong; for example, most NEOs have a \texttt{digest2} scores of zero in the Jtr and JFC classes, whereas many non-NEOs show nonzero scores there. Other categories, such as Hildas or Hansa, appear to be less discriminating.

In addition to the visual inspection, we attempted to identify optimal \textit{digest2} thresholds within each category to reduce non-NEOs while retaining genuine NEOs. That is, we sought to identify a boundary in each histogram (either above or below a given score) that would exclude zero, one, or two NEOs at most in our data sample, that is, a controlled tolerance for NEO losses. We divided our data samples into a training sample (2019 - 2023) and a validation sample (2024).

Tables~\ref{tab:filter0}, \ref{tab:filter1}, and \ref{tab:filter2} list the thresholds for three different filter modes ($filter_{0}$, $filter_{1}$, and $filter_{2}$), showing how many NEOs and non-NEOs each filter would remove. For example, $filter_{0}$ rejected 911 non-NEOs over five years without excluding any NEOs, whereas $filter_{2}$ eliminated 1,557 non-NEOs, but at the cost of 14 NEOs. We then applied these same filters to the 2024 data (Table~\ref{tab:filter2024}). Depending on the chosen tolerance, between 159 and 278 non-NEOs were selected from the NEOCP, and only three or four genuine NEOs out of 3,427 discovered that year were missed. Even though the filters do not classify the majority of the data (Tables~\ref{tab:confusion_matrix} and \ref{tab:confusion_matrix_2024}), they demonstrate exceptional precision in correctly identifying non-NEOs; it exceeds 98\%. This high precision suggests that basic filters can serve as an effective initial screening tool for filtering objects on the NEOCP. This significantly reduces the number of candidates that require further analysis.

\begin{table}[h!]
\centering
\caption{$Filter_{0}$. Threshold values and object counts for each \textit{digest2} type, showing the number of correctly identified non-NEOs and the maximum number of falsely rejected NEOs (0) in each category based on 2019–2023 NEOCP data.}
\begin{tabular}{l|c|c|c}
\textbf{Digest2 Type} & \textbf{Threshold} & \textbf{Non-NEOs} & \textbf{NEOs } \\
\midrule
Hun1   & >34 & 280 & 0 \\
Hun2   & >14 & 38  & 0 \\
JFC1   & >10 & 130 & 0 \\
JFC2  & >13 & 35  & 0 \\
JTr1   & >2  & 223 & 0 \\
JTr2   & >1  & 162 & 0 \\
MB1\_1 & >26 & 74  & 0 \\
MB1\_2 & >21 & 41  & 0 \\
MB2\_1 & >74 & 0   & 0 \\
MB2\_2 & >75 & 0   & 0 \\
MB3\_1 & >8  & 138 & 0 \\
MB3\_2 & >11 & 19  & 0 \\
Pal1   & >10 & 180 & 0 \\
Pal2   & >9  & 19  & 0 \\
\hline
Total & & 911 & 0\\
\end{tabular}
\label{tab:filter0}
\end{table}

\begin{table}[h!]
\centering
\caption{$Filter_{1}$. Threshold values and object counts for each \textit{digest2} type, showing the number of correctly identified non-NEOs and the maximum number of falsely rejected NEOs (1) in each category based on 2019–2023 NEOCP data.}
\begin{tabular}{l|c|c|c}
\textbf{Digest2 Type} & \textbf{Threshold} & \textbf{Non-NEOs} & \textbf{NEOs } \\
\midrule
Hun1   & >30 & 373 & 1 \\
Hun2   & >12 & 99  & 1 \\
JFC1   & >10 & 130 & 0 \\
JFC2  & >11 & 52  & 1 \\
JTr1   & >0  & 367 & 1 \\
JTr2   & >0  & 257 & 1 \\
MB1\_1 & >13 & 127  & 1 \\
MB1\_2 & >10 & 116  & 1 \\
MB2\_1 & >33 & 58   & 1 \\
MB2\_2 & >19 & 65   & 1 \\
MB3\_1 & >5  & 208 & 1\\
MB3\_2 & >3 & 107  & 1 \\
Pal1   & >8 & 255 & 1 \\
Pal2   & >4  & 131  & 1 \\
\hline
Total & & 1362 & 9\\
\end{tabular}
\label{tab:filter1}
\end{table}

\begin{table}[h!]
\centering
\caption{$Filter_{2}$. Threshold values and object counts for each \textit{digest2} type, showing the number of correctly identified non-NEOs and the maximum number of falsely rejected NEOs (2) in each category based on 2019–2023 NEOCP data.}
\begin{tabular}{l|c|c|c}
\textbf{Digest2 Type} & \textbf{Threshold} & \textbf{Non-NEOs} & \textbf{NEOs } \\
\midrule
Hun1   & >28 & 434 & 2 \\
Hun2   & >12 & 99  & 1 \\
JFC1   & >5 & 310 & 2 \\
JFC2  & >10 & 66  & 2 \\
JTr1   & >0  & 367 & 1 \\
JTr2   & >0  & 257 & 1 \\
MB1\_1 & >12 & 134  & 2 \\
MB1\_2 & >6 & 177  & 2 \\
MB2\_1 & >25 & 96   & 2 \\
MB2\_2 & >17 & 84   & 2 \\
MB3\_1 & >4  & 250 & 2\\
MB3\_2 & >2 & 197  & 2 \\
Pal1   & >7 & 301 & 2 \\
Pal2   & >4  & 131  & 1 \\
\hline
Total & & 1557 & 14\\
\end{tabular}
\label{tab:filter2}
\end{table}

\begin{table}[h!]
\centering
\caption{Arbitrary filters for individual \textit{digest2} thresholds for the 2024 dataset of NEOCP candidates.}
\begin{tabular}{l|c|c}
\textbf{Filter} & \textbf{non-NEOs} & \textbf{NEOs}  \\
\midrule
$Filter_{0}$   & 159 &  3 \\
$Filter_{2}$   &  252 &  4  \\
$Filter_{1}$   & 278 & 4 \\
\hline
Total  &1308  & 3,427\\
\end{tabular}
\tablefoot{The second and third column show excluded candidates per orbit class. The filter would eliminate hundred(s) of non-NEOs at the cost of losing three or four NEOs in 2024. The bottom column shows the total number of non-NEOs and NEOs discovered in 2024.}
\label{tab:filter2024}
\end{table}

\begin{table*}[ht]
\centering
\caption{Confusion matrices for filters 0–2 using 2019–2023 NEOCP data.}
\label{tab:confusion_matrix}
\begin{tabular}{l@{\hskip 2em}ccc@{\hskip 2em}ccc@{\hskip 2em}ccc}
\toprule
& \multicolumn{9}{c}{\textbf{Predicted Label}} \\
\cmidrule(lr){2-10}
\textbf{True Label}
& \multicolumn{3}{c}{\textbf{Filter 0}}
& \multicolumn{3}{c}{\textbf{Filter 1}}
& \multicolumn{3}{c}{\textbf{Filter 2}} \\
\cmidrule(lr){2-4} \cmidrule(lr){5-7} \cmidrule(lr){8-10}
& non-NEO & Unclassified & Total
& non-NEO & Unclassified & Total
& non-NEO & Unclassified & Total \\
\midrule
NEO      & 0    & 15591 & 15591
         & 9    & 15582 & 15591
         & 14   & 15577 & 15591 \\
non-NEO  & 911  & 7085  & 7996
         & 1362 & 6634  & 7996
         & 1557 & 6439  & 7996 \\
\bottomrule
\end{tabular}
\tablefoot{Most of the data are not classified, but the precision of classifying non-NEOs among the classified sample is very high ($\gg$99\%).}
\end{table*}

\begin{table*}[ht]
\centering
\caption{Confusion matrices for filters 0–2 using 2024 NEOCP data and filters derived from 2019–2023 data.}
\label{tab:confusion_matrix_2024}
\begin{tabular}{l@{\hskip 2em}ccc@{\hskip 2em}ccc@{\hskip 2em}ccc}
\toprule
& \multicolumn{9}{c}{\textbf{Predicted Label}} \\
\cmidrule(lr){2-10}
\textbf{True Label}
& \multicolumn{3}{c}{\textbf{Filter 0}}
& \multicolumn{3}{c}{\textbf{Filter 1}}
& \multicolumn{3}{c}{\textbf{Filter 2}} \\
\cmidrule(lr){2-4} \cmidrule(lr){5-7} \cmidrule(lr){8-10}
& non-NEO & Unclassified & Total
& non-NEO & Unclassified & Total
& non-NEO & Unclassified & Total \\
\midrule
NEO      & 3   & 3424 & 3427
         & 4   & 3423 & 3427
         & 4   & 3423 & 3427 \\
non-NEO  & 159 & 1149 & 1308
         & 252 & 1056 & 1308
         & 278 & 1030 & 1308 \\
\bottomrule
\end{tabular}
\tablefoot{Most of the data are not classified, but the precision of classifying non-NEOs among the classified sample is very high ($\gg$99\%).}
\end{table*}

\section{Machine-learning methods}
\label{sec:ML}

Although the refined \texttt{digest2} filters can reduce the number of non-NEOs on the NEOCP, these methods rely on a handful of fixed thresholds. Machine-learning (ML) approaches offer an alternative that can detect nonlinear relationships in high-dimensional data without depending on predefined cutoffs. By training on historical examples of NEO and non-NEO tracklets, ML models potentially classify objects more accurately and lessen the need for follow-up observations on false positives.

This study used observational records in the MPC1992 obs80-column format, which provides the right ascension (RA), the declination (Dec), the magnitude, the epoch, and the band for each tracklet. Crucially, our dataset consisted exclusively of known NEO and non-NEO tracklets, giving us reliable orbit classifications a priori. The training portion covered five years (2019--2023), comprising 15,591 NEO tracklets and 7,996 non-NEO tracklets, while all data for 2024, consisting of 4,737 tracklets (3,428 labeled as NEOs), were used as the test set for the final evaluation.

\subsection{Data preprocessing}
The data quality is critical for ML applications, as well as in astronomy, where measurements can be degraded by instrumental limitations and atmospheric conditions or other noise sources. To ensure data integrity, records containing missing values were systematically excluded using a complete case analysis on each row. This preprocessing step resulted in the removal of 6 samples from our training dataset, yielding a final count of 24,913 samples. The low proportion of missing data (0.02\%) suggests that the completeness of our observational dataset was high, and this minimizes any potential impact on the model training.

In addition to handling missing values, we addressed the class imbalance in our dataset. The initial distribution showed a significant skew, with 16,543 NEO tracklets versus 8,376 non-NEO tracklets. To mitigate potential classification bias, we implemented a random undersampling strategy that reduced the majority class (NEOs) to match the size of the minority class. This balanced-sampling approach resulted in a final training set of 8,375 tracklets per class, ensuring equal representation of NEOs and non-NEOs while maintaining the statistical properties of both populations. A random sampling was performed with a fixed seed value to ensure the reproducibility of our results.

\subsection{Model architecture and implementation}
We implemented four distinct ML methods so that we were able to examine a spectrum of modeling capabilities. First, we adopted a gradient-boosting machine (GBM) based on XGBoost using binary logistic regression as the objective function, a log-loss metric, 100 estimators, a learning rate of 0.1, and a maximum tree depth of 5. This approach sequentially corrected its own errors through boosting, which makes it well suited to complex nonlinear relations in the orbital parameter space. Second, we trained a random forest (RF) classifier using the scikit-learn framework. This algorithm builds multiple decision trees on randomized subsets of the training data and features and then aggregates their predictions. Random forest naturally handles outliers, reduces overfitting by averaging, and quantifies the contribution of each feature to the final classification.

As a simpler alternative, we also employed the code called stochastic gradient descent (SGD) configured as a linear classifier, using binary cross-entropy loss and an adaptive learning rate scheduling. Because it fits a single linear decision boundary, SGD offers computational efficiency and is relatively interpretable: The model coefficients directly indicate which orbital parameters carry the most weight for a classification. To account for any remaining imbalance, we enabled balanced class weights. Finally, we experimented with a feed-forward neural network (NN) containing two hidden layers of 64 ReLU-activated neurons apiece, binary cross-entropy as the loss function, and an Adam optimizer with a learning rate of 0.001. This setup aimed to capture complex relations through nonlinear activation, while validation-based early stopping prevented overfitting.

We evaluated each model performance using accuracy, precision, recall, and F1-score. These metrics were computed on a stratified validation split to preserve the relative frequencies of NEOs and non-NEOs. After we refined the hyperparameters on the validation subset, the models were tested on the 2024 dataset for an unbiased measure of real-world applicability. A correct classification of NEOs is crucial to ensure that follow-up observations are not missed, and the correct rejection of non-NEOs helps us to save telescope time.

To take advantage of the strengths of individual classifiers, we investigated two ensemble strategies. First, we applied a model stacking, where the output of GBM, RF, SGD, and NN was combined by a logistic regression meta-learner, thus capturing the complementary aspects of the predictions of each algorithm. Second, we considered a voting mechanism that aggregated predictions either by majority rule (hard voting) or by averaging class probabilities (soft voting). We tuned the voting weights on the validation data to balance precision and recall. These ensemble methods were designed to mitigate the weaknesses of any single model, with the aim of achieving a more robust overall classification. By comparing threshold-based \texttt{digest2} filtering and multiple ML classifiers, we hoped to identify the most promising routes to reduce false positives on the NEOCP while retaining the highest fraction of genuine NEOs.

\section{Results}
\label{sec:results}

\begin{table}[ht]
\centering
\small % Reduces font size for a smaller table
\caption{Model performance metrics}
\begin{tabular}{|c|c|c|c|c|}
\hline
\multicolumn{2}{|c|}{\textbf{GBM}} & \multicolumn{3}{c|}{\textbf{Accuracy=0.9265}} \\ \hline
\textbf{Metric}       & \textbf{Precision} & \textbf{Recall} & \textbf{F1-score} & \textbf{Support} \\ \hline
\textbf{NEO}      & 0.98               & 0.92            & 0.95              & 3428             \\ \hline
\textbf{Non-NEO}      & 0.82               & 0.95            & 0.88              & 1308             \\ \hline
\textbf{Macro Avg}    & 0.90               & 0.93            & 0.91              & 4736             \\ \hline
\textbf{Weighted Avg} & 0.93               & 0.93            & 0.93              & 4736             \\ \hline\hline

\multicolumn{2}{|c|}{\textbf{RF}} & \multicolumn{3}{c|}{\textbf{Accuracy=0.9261}} \\ \hline
\textbf{NEO}      & 0.98               & 0.92            & 0.95              & 3428             \\ \hline
\textbf{Non-NEO}      & 0.81               & 0.95            & 0.88              & 1308             \\ \hline
\textbf{Macro Avg}    & 0.90               & 0.93            & 0.91              & 4736             \\ \hline
\textbf{Weighted Avg} & 0.93               & 0.93            & 0.93              & 4736             \\ \hline\hline

\multicolumn{2}{|c|}{\textbf{SGD}} & \multicolumn{3}{c|}{\textbf{Accuracy=0.9269}} \\ \hline
\textbf{NEO}      & 0.95               & 0.94            & 0.95              & 3428             \\ \hline
\textbf{Non-NEO}      & 0.86               & 0.88            & 0.87              & 1308             \\ \hline
\textbf{Macro Avg}    & 0.91               & 0.91            & 0.91              & 4736             \\ \hline
\textbf{Weighted Avg} & 0.93               & 0.93            & 0.93              & 4736             \\ \hline\hline

\multicolumn{2}{|c|}{\textbf{NN}} & \multicolumn{3}{c|}{\textbf{Accuracy=0.9329}} \\ \hline
\textbf{NEO}      & 0.98               & 0.93            & 0.95              & 3428             \\ \hline
\textbf{Non-NEO}      & 0.84               & 0.94            & 0.89              & 1308             \\ \hline
\textbf{Macro Avg}    & 0.91               & 0.93            & 0.92              & 4736             \\ \hline
\textbf{Weighted Avg} & 0.94               & 0.93            & 0.93              & 4736             \\ \hline
\end{tabular}
\label{tab:model_metrics}
\end{table}

\begin{figure}[ht]
   \includegraphics[width=\linewidth]{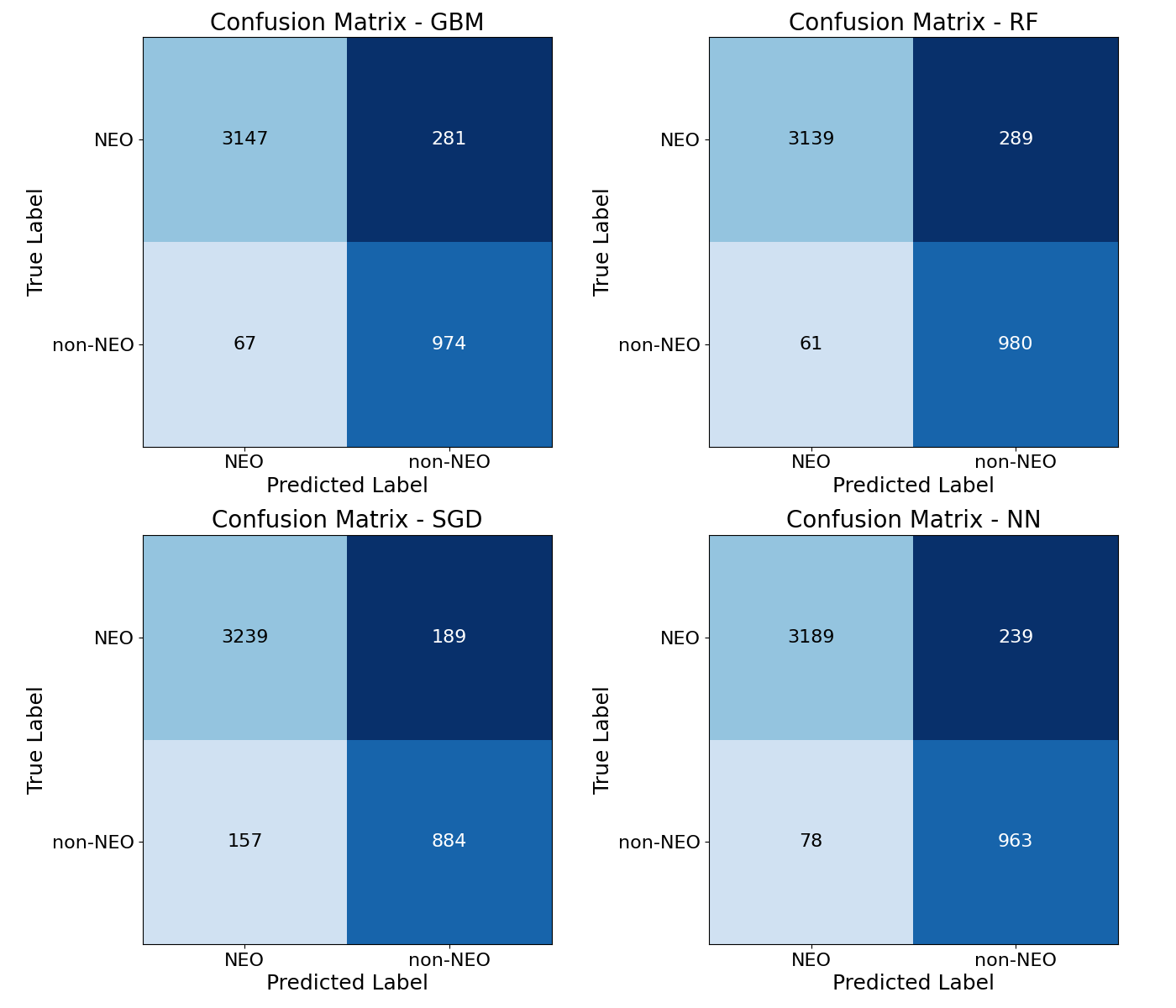}
    \caption{Comparison of the four ML methods we used on 2024 NEOCP data. The SGD method performs best. It misclassified 189 NEOs as non-NEOs. The RF misclassified the most NEOs as non-NEOs.}
    \label{fig:confusion_matrix}
\end{figure}

Four machine-learning (ML) models (GBM, RF, SGD, and NN) were trained on NEOCP discovery tracklets (2019–2023) and tested on an independent dataset from 2024. Table~\ref{tab:model_metrics} summarizes the overall performance of each method in the 2024 test dataset. The results demonstrate that the four methods achieve a broadly comparable performance, with accuracies ranging from 0.9261 (RF) to 0.9329 (NN). The precision is highest for the identification of NEOs (ranging from 0.95 to 0.98) and lower for non-NEOs (0.81 to 0.86). Despite the similarity of these metrics, the SGD classifier was chosen for the detailed analysis because it produced the fewest misidentified NEOs in the test set. Figure~\ref{fig:confusion_matrix} shows the confusion matrices for all four methods. In the SGD matrix, the upper left quadrant represents correctly identified NEOs, the lower right quadrant shows correctly identified non-NEOs, and the off-diagonal cells represent misclassifications. The SGD classifier incorrectly labeled 189 NEOs as non-NEOs and correctly classified 3,239 NEOs, along with 884 non-NEOs.

\begin{figure}[ht]
   \includegraphics[width=\linewidth]{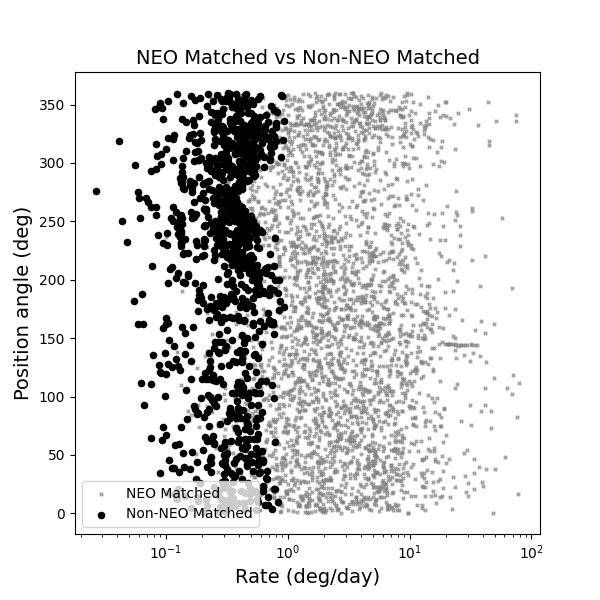}
   \includegraphics[width=\linewidth]{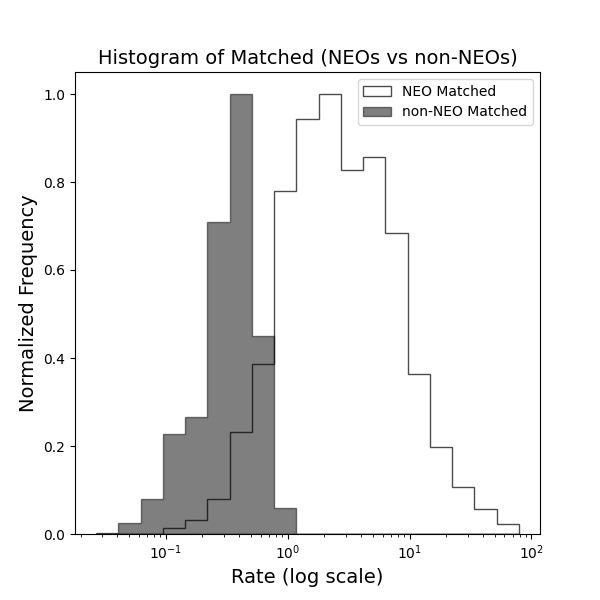}
    \caption{Rate of motion and position angle (top) and histogram of rate of motion (bottom) for correctly identified NEOs and non-NEOs by SGD. There is a clear distinction between NEOs and non-NEOs in rate of motion.}
    \label{fig:rates_matched}
\end{figure}

\begin{figure}[ht]
   \includegraphics[width=\linewidth]{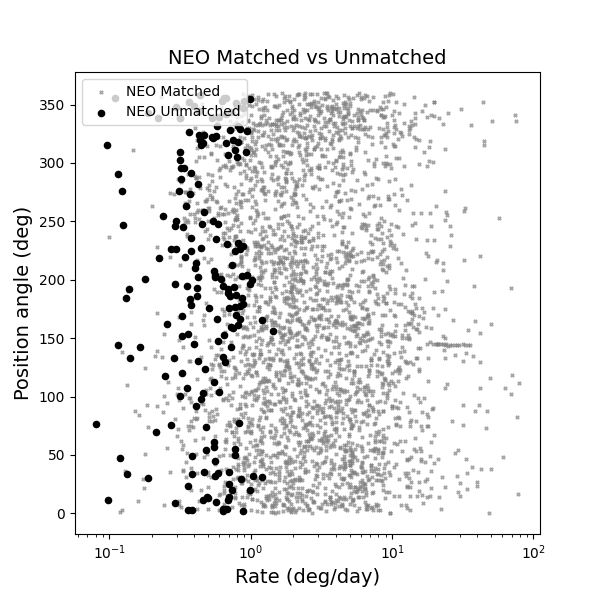}
   \includegraphics[width=\linewidth]{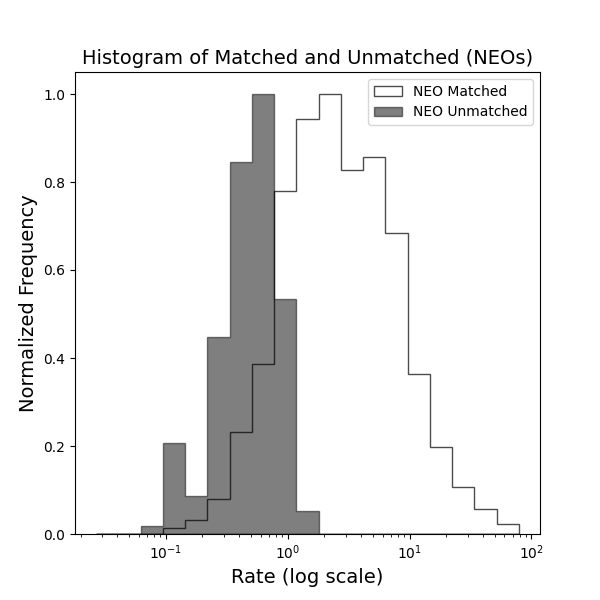}
    \caption{Rate of motion and position angle (top) and histogram of rate of motion (bottom) for correctly and incorrectly identified NEOs identified by SGD. Misidentified NEOs are slower than the correctly identified ones.}
    \label{fig:rates_neos}
\end{figure}

\begin{figure}[ht]
   \includegraphics[width=\linewidth]{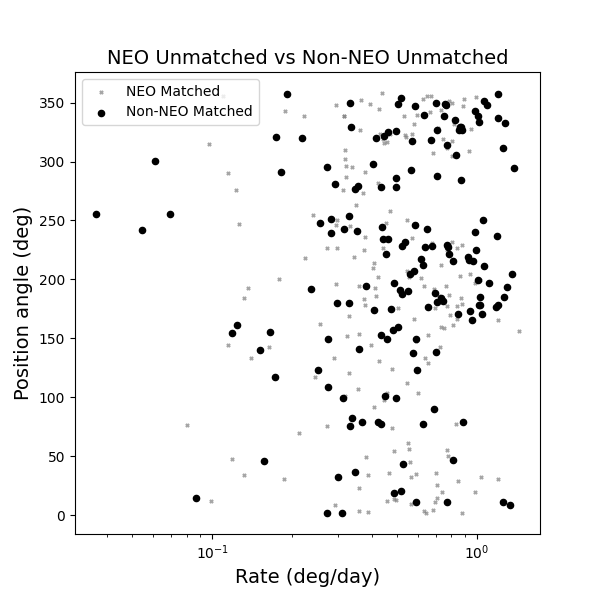}
   \includegraphics[width=\linewidth]{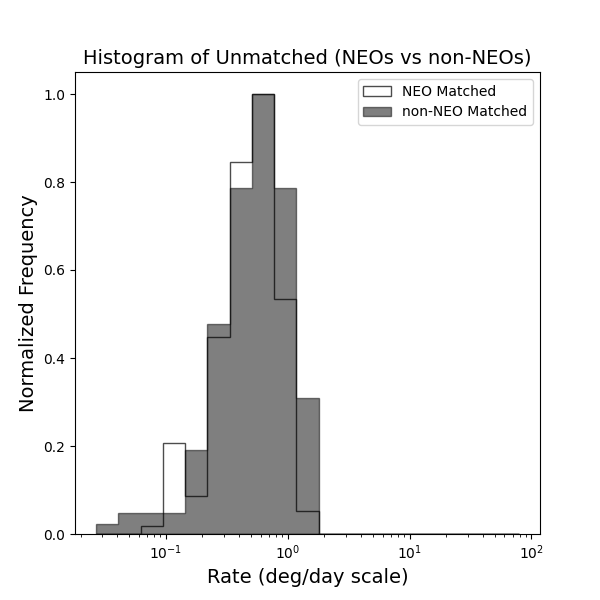}
    \caption{Rate of motion and position angle (top) and histogram of rate of motion (bottom) for incorrectly identified NEOs and non-NEOs by SGD. The motion properties of the two orbital types are similar.}
    \label{fig:rates_unmatches}
\end{figure}

\begin{figure}[ht]
   \includegraphics[width=\linewidth]{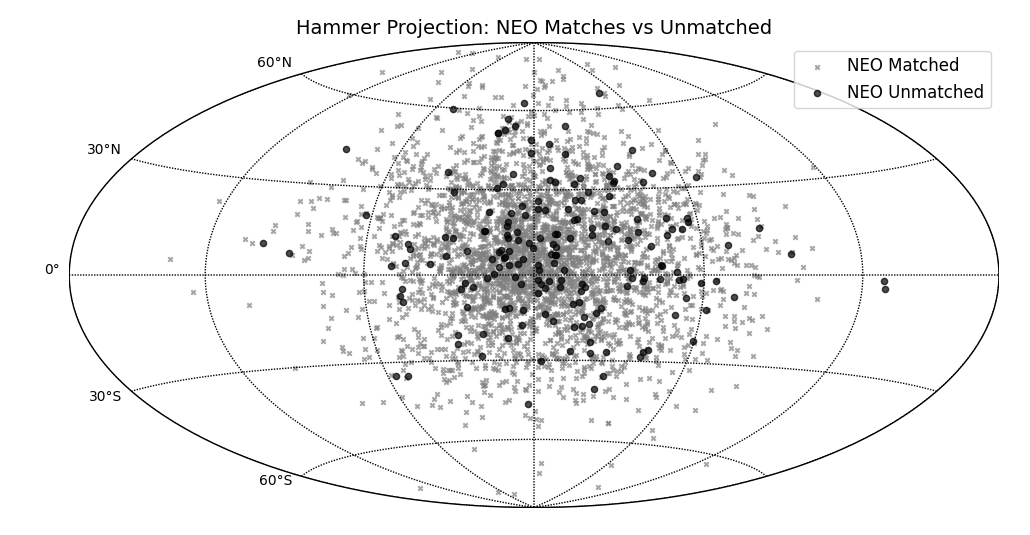}
    \caption{Opposition-centric ecliptical coordinates of identified and misidentified NEOs by SGD.}
    \label{fig:hammer}
\end{figure}

We investigated the 189 misidentified NEOs in detail. Figures~\ref{fig:rates_matched}, \ref{fig:rates_neos}, and \ref{fig:rates_unmatches} show the apparent rates of motion and position angle of the motion vector for the matched and unmatched NEOs and non-NEOs from the SGD model. These figures reveal that slow-moving NEOs are more likely to be misclassified, while the orientation of their motion vectors has little impact.
We also investigated the sky-plane distribution of correctly and incorrectly classified objects. Figure~\ref{fig:hammer} shows that the opposition-centric ecliptical coordinates do not differentiate the two groups. Thus, incorrect classifications appear to be primarily related to on-sky motion rates and not to location.

\subsection{Decreasing the false-positive rate}
\begin{figure}[ht]
   \includegraphics[width=0.23\textwidth]{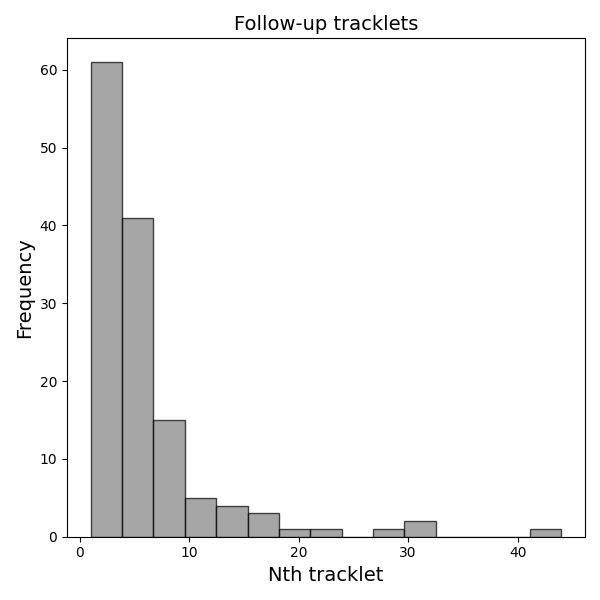}
      \includegraphics[width=0.23\textwidth]{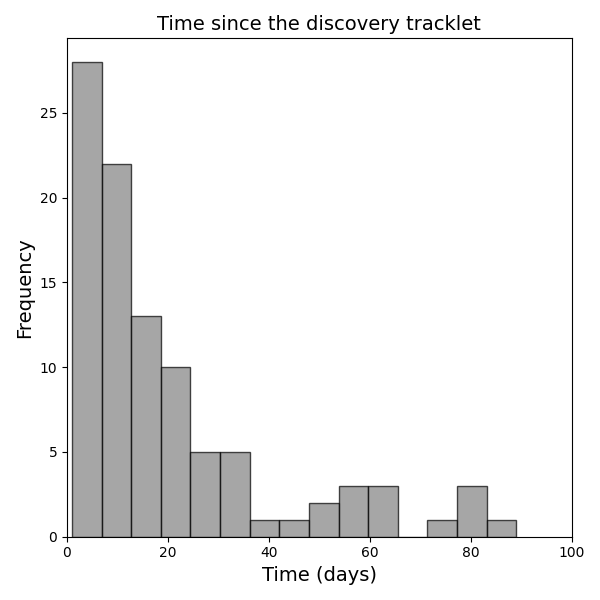}

    \caption{Histogram of the order of the first follow-up tracklet of initially misclassified NEOs that the ML model identified correctly (left). We also show the histogram of the time that elapsed since the original discovery tracklet (right) until the follow-up tracklet is classified correctly as a NEO for the first time.}
    \label{fig:delta_t}
\end{figure}

Although a misclassification of 189 of the 3428 NEOs (for 2024) may initially seem significant, it is important to note that our analysis was only applied to the discovery tracklets of objects when they first appeared on the NEOCP. Our previous work~\citep{keys2019digest2} demonstrated that the NEO \textit{digest2} score changes as a function of time and observational geometry, implying that many of these initially missed NEOs could be reidentified when additional data become available.
To examine this effect, all follow-up tracklets from the 189 misidentified NEOs were reevaluated using the SGD classifier. Of the 189 objects, 140 were correctly identified as NEOs in at least one follow-up tracklet. As shown in Figure~\ref{fig:delta_t}, most were correctly reclassified within a few days after the initial discovery tracklet. However, a small number required up to three months to receive a definitive NEO classification.

Four of the remaining 49 NEOs that were consistently misclassified in all available follow-up tracklets were 
already known (e.g., recoveries), leaving 45 new NEO misidentifications.
Table~\ref{tab:rejected_NEOs_ML} lists the original trksub, packed designation, and basic orbital elements and absolute magnitude (H) of these 45 objects. Most are relatively large and distant. They belong to Amor-class asteroids and have a low NEO noid \textit{digest2} score. None qualifies as a potentially hazardous asteroid (PHA; defined as having a minimum orbit intersection distance <0.05 AU and $H<22$) or as imminent Earth impactors 
\footnote{K19M00O,K22E05B,K22W01J,K23C01X,K24B01X,K24R01W}. Furthermore, since most were discovered toward the end of 2024, their follow-up arcs are very short. We speculate that with additional follow-up observations, many of these objects might eventually be correctly classified by our model and subsequently reach the NEOCP. 

\subsection{Follow-up tracklets of misidentified NEOs by filtering.}

In the 2024 NEOCP dataset, the basic filters misclassified only a few NEOs (three or four) while correctly labeling between 159 and 287 non-NEOs, depending on the filter that was applied. We further examined all 18 NEOs (from 2019–2024) that were misclassified by $filter_{2}$ and analyzed their follow-up tracklets using the same filter. We found that each of these 18 NEOs had at least one follow-up tracklet that was correctly classified. This was typically by the third follow-up. Additionally, when the original 18 misclassified discovery tracklets were processed through the SGD ML method, they were all misclassified again, indicating a strong correlation between the basic filtering and the SGD ML approach regarding misidentified NEOs. In contrast, the candidates classified as non-NEOs by $filter_{2}$ were consistently confirmed as non-NEOs by the SGD ML method.

\subsection{Analysis of unconfirmed candidates}

We analyzed 3725 unconfirmed NEOCP candidates (2019–2024) using the SGD ML method and the basic \textit{digest2} filter ($filter_{2}$). The unconfirmed data were divided into two subsets: 3190 candidates that remained unconfirmed, and 535 candidates that were later attributed to known objects (and thus have known orbital properties).

Within the attributed subset, the SGD ML method achieved a precision of 91\%, with approximately 50\% of the candidates classified as NEOs. The \texttt{filter2} correctly identified 79 unconfirmed candidates as non-NEOs, without eliminating any true NEOs. In contrast, when applied to the entire unconfirmed dataset, the SGD ML method indicated that 2620 candidates could be NEOs and only 570 were non-NEOs (see Table~\ref{tab:lost_neos}). This suggests that approximately 83\% of the unconfirmed NEOCP candidates are likely NEOs. The observed discrepancy in the NEO/non-NEO ratio of the unconfirmed and attributed subsets can be attributed to a selection effect: Non-NEOs are generally easier to confirm through follow-up observations via linking, which leads to their preferential attribution. $filter_{2}$ would have eliminated 100 unconfirmed candidates as non-NEOs.

\begin{table}[h!]
\caption{Unconfirmed NEO candidates: SGD ML classification (middle) and already attributed candidates (right) that were initially unconfirmed.}
\centering
\begin{tabular}{l|c|c}
\hline
Orbit type & Count - unconfirmed & Count - attributed \\
\hline
NEO & 2620 & 255\\
Non-NEO & 570 & 280 \\
\hline
\end{tabular}
\label{tab:lost_neos}
\end{table}

\begin{figure}[ht]
   \includegraphics[width=0.23\textwidth]{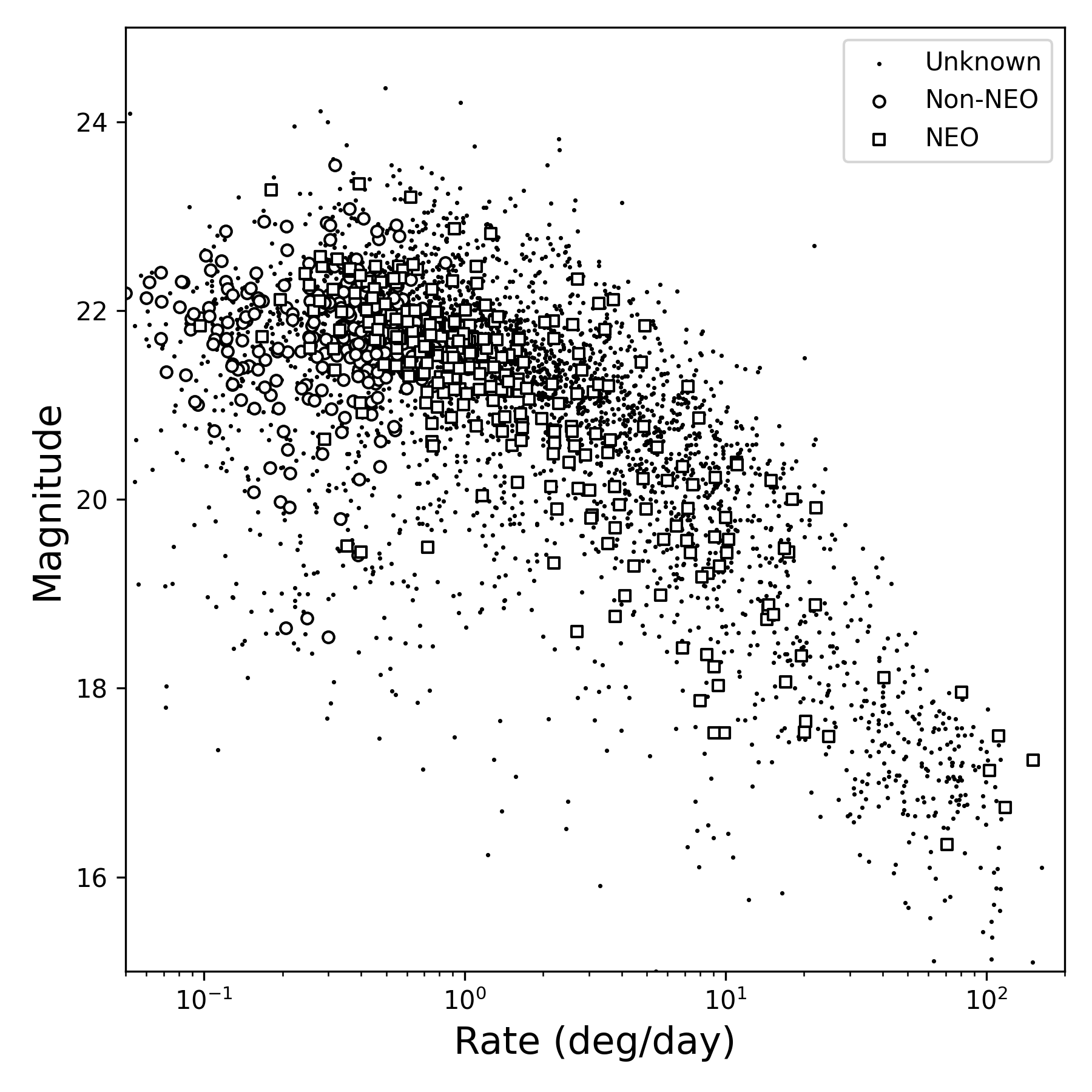}
      \includegraphics[width=0.23\textwidth]{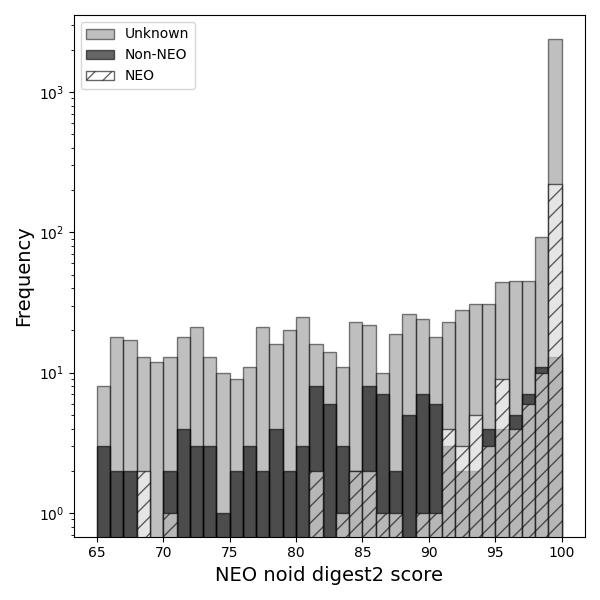}

    \caption{Rate of motion and mean V-band magnitude of unconfirmed NEO candidates. We highlight NEOs and non-NEOs that were attributed later (left). The histogram shows the NEO noid \textit{digest2} score of unconfirmed and attributed NEOCP candidates.}
    \label{fig:lost_data}
\end{figure}

\begin{figure}[ht]
   \includegraphics[width=0.23\textwidth]{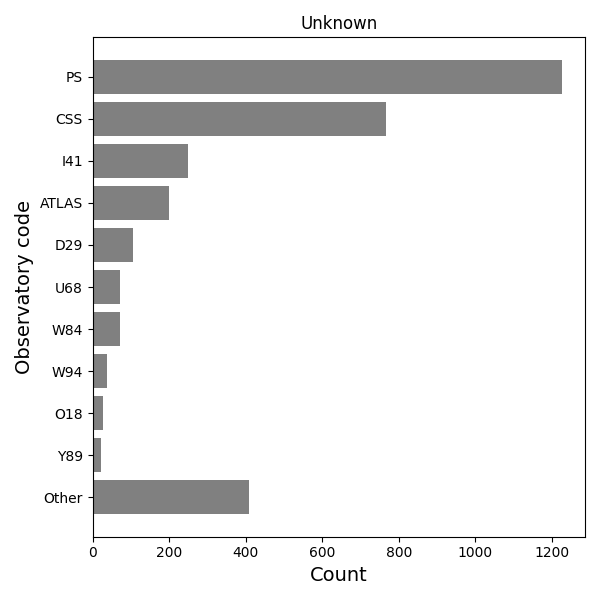}
      \includegraphics[width=0.23\textwidth]{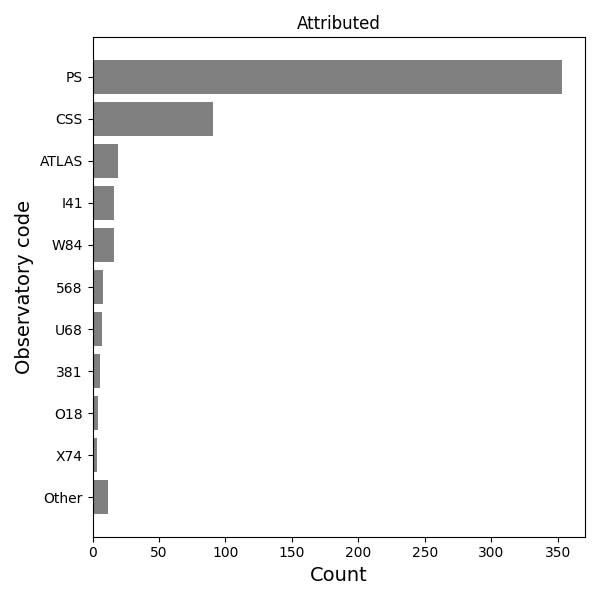}

    \caption{Observatory codes of tracklets that remain unconfirmed (left) and those that were later attributed (right).}
    \label{fig:lost_obscodes}
\end{figure}

Figure~\ref{fig:lost_data} illustrates the relation between the apparent magnitude and rate of motion of the unconfirmed candidates, together with the corresponding NEO noid \textit{digest2} scores. It is apparent that the unconfirmed objects generally fall into two categories: those moving extremely fast on the sky, which results in rapidly increasing positional uncertainties that hinder a timely recovery, and those that are very faint, making them difficult to detect with follow-up telescopes. Furthermore, as noted in the left panel of Figure~\ref{fig:lost_obscodes}, the distribution of observatory codes for tracklets that remain unconfirmed differs from that of attributed tracklets (Figure~\ref{fig:lost_obscodes}, right), reflecting differences in survey coverage and follow-up capability.

\section{Discussion}
\label{sec:discussion}

For the past 15 years, a single threshold of NEO noid \textit{digest2} (>65) has been used to select candidates, with a brief experiment in mid-2010 where the threshold was lowered in an attempt to identify more NEOs. However, this change resulted in an overwhelming number of non-NEO candidates without a significant increase in genuine NEO discoveries. As shown in Figure~\ref{fig:main_hist}, the number of NEOs with low \textit{digest2} scores is very small. Between 2019 and 2023, only 37 NEOs were found in the \textit{digest2} score range of 65 to 70, compared to 826 non-NEOs. In other words, the number of low-score NEOs is lower by an order of magnitude than the annual number of unconfirmed NEOCP candidates. Consequently, our first recommendation is to consider increasing the NEO noid \textit{digest2} threshold by a few points.

Our second recommendation is to leverage the full range of \textit{digest2} parameters and thresholds derived in Section ~\ref{sec:digest2 filtering}. By carefully tuning these thresholds, it is possible to eliminate a substantial fraction of non-NEO candidates from the NEOCP while keeping the misclassification of actual NEOs to a minimum. For example, a conservative $filter_{0}$ would eliminate no NEOs, while a more aggressive filter $filter_{2}$ would reject 14 NEOs over a five-year period, removing approximately 20\% of non-NEOs. When we consider that approximately 3,000 NEOs are discovered annually and around 600 objects remain unconfirmed, the misclassification of a few NEOs per year appears to be an acceptable compromise.

It is also important to note that more than a dozen NEOs are discovered each year through the lineage of isolated tracklets: the tracklets that never achieve a \textit{digest2} score of 65, but still belong to previously unknown NEOs. The Minor Planet Center (MPC) credits these discoveries in MPEC publications, often with a note such as “The following astrometry was found among the isolated tracklets and linked by R. Weryk.” For example, in 2024, 25 discoveries like this were announced (e.g., \citep{2024MPEC}).

A detailed analysis of the 14 rejected NEOs in the sample 2019–2023, together with 4 NEOs from 2024 (see Table~\ref{tab:rejected_NEOs_filter}), revealed that four objects lacked a NEO noid \textit{digest2} score exceeding 65 and therefore had to be manually added to the NEOCP. The remaining objects exhibited relatively low \textit{digest2} scores and a slow to moderate rate of motion typical of newly discovered NEOs. Of these, nine are of the Amor type (usually more distant objects), seven are Apollo, and two are Atira. In particular, all four machine-learning methods consistently identified the objects rejected by the filtering process as NEOs. This suggests that a combined approach using both machine-learning and traditional filtering techniques could offer significant benefits in reducing the non-NEO load on the NEOCP.

\begin{table}[h!]
\caption{NEOs that were rejected by \textit{digest2} filtering by $filter_{3}$.}
\centering
\begin{tabular}{l|c|c|c|c|c|c|c|c}
\hline
 \textbf{trksub} & \textbf{d2} & \textbf{stn} & \textbf{mag} & \textbf{desig} & \textbf{H} & \textbf{type} \\
\hline
P10RD4L & 88 & F51 & 21.9 & K19S10J & 17.5 & Amor  \\
C14GHT2 & 91 & G96 & 20.9 & K19T07F & 17.9 & Amor \\
P20TIfj & 90 & F52 & 19.5 & K19U11B & 20.6 & Amor  \\
P1152JZ & 90 & F51 & 18.9 & K20O08C & 20.8 & Amor  \\
\textbf{C5EA9Z2} & 17 & G96 & 20.7 & K21G06Q& 20.8& Apollo\\
\textbf{C59N5U2} & 13 & G96 & 19.7 & K21G03J & 23.7 & Apollo \\
P11gQVB & 67 & F51 & 22.8 & K21L03V & 24.0 & Amor  \\
\textbf{vk08c06} & 7 & T14 & 21.8 & K21L04J & 20.1 & Atira  \\
v13aug1 & 79 & 807 & 19.3 & K21P27H & 17.7 & Atira  \\
P21ltbf & 67 & F52 & 22.3 & K21S05C & 22.7 & Apollo  \\
P11BfaH & 83 & F51 & 20.4 & K22T14E & 22.3 &Apollo \\
C0E4805 & 87 & V00 & 22.1 & K22W05C & 25.4 &Amor \\
P21J2QE & 89 & F52 & 21.2 & K23O24E & 20.9 & Amor  \\
P11Nmhg & 75 & F51 & 21.5 & K23V03G & 22.6 & Apollo  \\
\hline
P21Vmx7 & 75 & F52 & 21.0 & K24L02Y & 22.8 & Apollo \\
\textbf{P21XFw4} & 63 & F52 & 21.7 & K24Q05E & 19.4 & Apollo\\
P222up9 & 66 & F52 & 21.5 & K24W05Z &21.1 & Amor \\
\textbf{P223svM} & 96& F51  & 20.9 &K24X15J  & 18.6& Amor \\
\hline
\end{tabular}
\tablefoot{Submitter trksub, NEO noid \textit{digest2} score, observatory code, absolute magnitude (H), packed designation, and orbit type. The trksubs in bold had to be added to NEOCP manually because of a low \textit{digest2} score. }
\label{tab:rejected_NEOs_filter}
\end{table}

Implementing $filter_{2}$ would remove approximately 19\% of non-NEOs from the NEOCP for the 2019–2023 period and 21\% for 2024. This reduction enables follow-up observers to focus on genuine NEOs instead of expending resources on candidates that are ultimately non-NEOs. Consequently, more NEOs were confirmed than unconfirmed objects.

All four machine-learning methods we evaluated achieved a similar precision and accuracy. We selected the Stochastic Gradient Descent (SGD) classifier as the most appropriate, however. The SGD method correctly labeled 95\% of the NEOs and 85\% of the non-NEOs. In particular, SGD is capable of reclassifying follow-up tracklets of initially misclassified NEOs as true NEOs, and only about 1\% of the NEOs remained misclassified. This high performance is promising, particularly since more than 80\% of non-NEOs are correctly identified and therefore excluded from the NEOCP. For example, in 2024, SGD would correctly classify 884 non-NEOs, which would prevent them from being posted to the NEOCP and would save significant follow-up effort. As shown in Table~\ref{tab:MBA_followup}, these non-NEOs correspond to 13,030 pointings on 4,150 tracklets from 104 observatories. Although nearly half of these pointings were incidental (e.g., collected by large-survey telescopes such as Pan-STARRS, the Catalina Sky Survey, and DECAM), the remaining detections represent thousands of valuable exposures that extend the observational arcs of non-NEOs.

With approximately 600 unconfirmed candidates per year and SGD suggesting that roughly half are true NEOs, eliminating non-NEOs from the NEOCP would allow follow-up observers to focus their resources on the most promising candidates. This refined focus would increase the likelihood that genuine NEOs are confirmed instead of expending effort on objects that are eventually ruled out by machine-learning or filtering.

Although modern surveys such as the VRO and NEO Surveyor promise alternative discovery methods, such as inter-night linking~\citep{Veres17b,Eggl20,Heinze22}, the NEOCP will likely remain in use by current and future survey teams. These new surveys may significantly alter the NEOCP population (e.g., by introducing many very faint objects), and the underlying \textit{digest2} score population model might change accordingly. However, the ML and filtering models can both be retrained rapidly to establish new thresholds. This ensures a continued efficiency in the selection of NEO candidates.

\begin{table}[h!]
\caption{Follow-up tracklets of non-NEOs that were correctly classified by SGD ML in 2024.}
\centering
\small
\begin{tabular}{|l|c|c}
\hline
pointings&13,030\\
\hline
tracklets&4,150\\
\hline
observatories&104\\
\hline
\end{tabular}
\tablefoot{About half of the pointings were submitted by surveys (Pan-STARRS, CSS, and DECAM) and were therefore incidental.}
\label{tab:MBA_followup}
\end{table}

\begin{table}[h!]
\caption{NEOs rejected by ML, including on all follow-up tracklets. }
\centering
\footnotesize
\begin{tabular}{|l|c|c|c|c|c|c|c|c}
\hline
 \textbf{trksub} & \textbf{d2} & \textbf{stn} & \textbf{mag} & \textbf{desig} & \textbf{H} & \textbf{type} \\
\hline

P11Qlp4&84&F51&21.81&K24A00Z&23.45&Amor\\
C0Q1LV5&81&V00&22.85&K24B01K&24.49&Amor\\
P11SWZO&86&F51&22.00&K24F05M&22.86&Amor\\
W016875&69&O18&21.40&K24F06G&22.46&Amor\\
CAFA6M2&81&G96&21.61&K24G04H&22.31&Amor\\
P21TNan&75&F52&21.92&K24H05W&23.39&Apollo\\
P21TZ8V&82&F52&21.94&K24J00J&25.61&Apollo\\
CANNPM2&92&G96&20.57&K24L01S&21.28&Amor\\
P21VXeH&93&F52&21.43&K24M02L&22.06&Amor\\
P11W6vo&91&F51&22.66&K24N01H&23.09&Amor\\
P11XzKQ&98&F51&20.77&K24O13J&20.03&Amor\\
P21X5sn&81&F52&22.23&K24P04T&22.96&Amor\\
P11XwR1&66&F51&21.13&K24Q00R&20.87&Amor\\
P21XzJF&75&F52&21.67&K24Q00Z&24.53&Amor\\
P11XD9B&79&F51&21.63&K24Q02N&23.41&Amor\\
P11XSlf&66&F51&21.62&K24R04M&21.34&Amor\\
P21XZNY&70&F52&21.14&K24R04U&21.07&Amor\\
P21Yd4E&92&F52&22.29&K24R09Q&25.27&Amor\\
P21Yet0&76&F52&22.53&K24R10X&23.93&Amor\\
P11XKzC&89&F51&22.12&K24R15E&22.10&Amor\\
P21YnHz&86&F52&22.77&K24R16F&24.78&Amor\\
P21YFpC&89&F52&21.35&K24R16U&22.02&Amor\\
P21YQz2&86&F52&21.69&K24R31C&26.85&Apollo\\
P11Zu8r&89&F51&22.66&K24S05M&23.44&Amor\\
P21ZzU1&89&F52&22.05&K24S06E&23.97&Amor\\
CAZYHF2&79&G96&21.98&K24T04G&22.52&Amor\\
CAYM7Z2&83&G96&22.17&K24T04Z&22.93&Amor\\
P120y9N&88&F51&22.21&K24T14U&23.37&Amor\\
CC36UF2&79&G96&20.33&K24U01V&22.10&Apollo\\
C0U87Q5&82&V00&22.38&K24U05F&21.37&Apollo\\
P121pLX&81&F51&21.35&K24U11K&23.88&Amor\\
P222up9&67&F52&21.46&K24W05Z&21.13&Amor\\
P222Jzr&94&F52&22.76&K24W14S&23.92&Amor\\
C0W5FY5&72&V00&21.99&K24X02D&24.86&Amor\\
C0WN425&95&V00&21.70&K24X04T&24.17&Amor\\
P123cKy&87&F51&21.64&K24X07S&21.41&Amor\\
P223mQp&85&F52&21.53&K24X15K&19.53&Apollo\\
MHD0712&89&W16&19.50&K24X16Q&17.60&Amor\\
P223e3m&88&F52&22.23&K24X18O&21.28&Amor\\
CCJ1PK2&90&G96&20.49&K24Y02C&21.86&Amor\\
C0YTF95&88&V00&21.84&K24Y03A&23.24&Amor\\
P124saO&93&F51&21.78&K24Y05Z&21.75&Amor\\
C106LP5&86&V00&21.77&K24Y10W&23.59&Amor\\
P124O24&80&F51&22.12&K24Y13C&19.77&Apollo\\
P224lzV&90&F52&21.59&K24Y14E&20.39&Amor\\
\hline
\end{tabular}
\tablefoot{Submitter trksub, NEO noid \textit{digest2} score, observatory code, absolute magnitude (H), packed designation, and orbit type.}
\label{tab:rejected_NEOs_ML}
\end{table}

\section{Data availability}
\label{sec:data_and_code}

The GitHub repository\footnote{\url{https://github.com/Smithsonian/ML_digest2_methods}} provides the following data and tools:
\begin{itemize}
    \item \textbf{data directory}: Contains the positions of tracklets in MPC1992 format (.obs extension) for NEOCP data from 2019 to 2024, computed \textit{digest2} values (split into 2019--2023 and 2024) in CSV format, the \texttt{MPC.config} configuration file for \textit{digest2}, optimal thresholds for basic filtering (JSON files), and lists of non-NEOs identified through basic filtering (\texttt{filtered\_pass.csv}).
    
    \item \textbf{src directory}: Contains Python executables:
    \texttt{find\_filter.py} (derives a basic filter from the input data and outputs a JSON file),
    \texttt{neocp\_filter.py} (applies a derived JSON filter to select non-NEOs from an input CSV file),
    \texttt{testing\_pipeline\_formatted.py} (uses trained machine learning models from the \texttt{models} directory to classify tracklets as NEOs or non-NEOs),
    and \texttt{training\_pipeline\_formatted.py} (trains four distinct machine learning models on an input sample).
    
    \item \textbf{models directory}: Contains the machine learning models developed in this work, intended for use with the \texttt{testing\_pipeline\_formatted.py} code.
\end{itemize}

\begin{acknowledgements}
This work was supported by the MPC's NASA cooperation agreement funding. We also acknowledge support of Oumuamua-Laukien fellowship awarded to the Galileo Project at Harvard University by the Laukien Science Foundation. 
\end{acknowledgements}

\bibliographystyle{aa}
\bibliography{references}

\end{document}